\documentclass[12pt,a4paper]{article}

\usepackage{amsmath}

\usepackage{xspace}

\usepackage{epsfig}

\usepackage{rotating}

\usepackage{dcolumn}

\usepackage{url}

\usepackage{hyperref}

\usepackage{ifthen} 
\usepackage{subfigure} 
\usepackage{rotating}

\usepackage{lineno}  
\usepackage{graphicx}  

\usepackage{xspace}  
\usepackage{color}
\usepackage{colortbl}

\newboolean{pdflatex}
\setboolean{pdflatex}{true} 
\newboolean{uprightparticles}
\setboolean{uprightparticles}{false} 
\usepackage{amssymb}
\usepackage{amsfonts}
\usepackage{upgreek}
\usepackage{rotating}
\usepackage{multirow}

\hypersetup{ pdfborder=0 0 0, urlbordercolor=1 0 0 }

\usepackage{cite}

\usepackage{graphicx}

\usepackage{color}

\usepackage{colortbl}

\def\lhcb {LHCb\xspace}

\ifthenelse{\boolean{uprightparticles}}%
{

 \def\Pmu         {\ensuremath{\upmu}\xspace}

 \def\Ppi         {\ensuremath{\uppi}\xspace}

 \def\Ppsi        {\ensuremath{\uppsi}\xspace}

 \def\PDelta      {\ensuremath{\Delta}\xspace}                 
 \def\PXi      {\ensuremath{\Xi}\xspace}                 
 \def\PLambda      {\ensuremath{\Lambda}\xspace}                 
 \def\PSigma      {\ensuremath{\Sigma}\xspace}                 
 \def\POmega      {\ensuremath{\Omega}\xspace}                 
 \def\PUpsilon      {\ensuremath{\Upsilon}\xspace}                 
 
 \def\PB      {\ensuremath{\mathrm{B}}\xspace}                 
                  
 \def\PD      {\ensuremath{\mathrm{D}}\xspace}

 \def\PJ      {\ensuremath{\mathrm{J}}\xspace}                 
 \def\PK      {\ensuremath{\mathrm{K}}\xspace}

 \def\PZ      {\ensuremath{\mathrm{Z}}\xspace}

 \def\Pc      {\ensuremath{\mathrm{c}}\xspace}

 \def\Pi      {\ensuremath{\mathrm{i}}\xspace}

}
{

 \def\Pmu         {\ensuremath{\mu}\xspace}

 \def\Ppi         {\ensuremath{\pi}\xspace}

 \def\Ppsi        {\ensuremath{\psi}\xspace}                 
                  
 \mathchardef\PDelta="7101
 \mathchardef\PXi="7104
 \mathchardef\PLambda="7103
 \mathchardef\PSigma="7106
 \mathchardef\POmega="710A
 \mathchardef\PUpsilon="7107
                  
 \def\PB      {\ensuremath{B}\xspace}                 
                  
 \def\PD      {\ensuremath{D}\xspace}

 \def\PJ      {\ensuremath{J}\xspace}                 
 \def\PK      {\ensuremath{K}\xspace}

 \def\PZ      {\ensuremath{Z}\xspace}

 \def\Pc      {\ensuremath{c}\xspace}

 \def\Pi      {\ensuremath{i}\xspace}

}

\def\mup        {\ensuremath{\Pmu^+}\xspace}
\def\mun        {\ensuremath{\Pmu^-}\xspace} 

\def\Z      {\ensuremath{\PZ^0}\xspace}

\def\c     {\ensuremath{\Pc}\xspace}

\def\pion  {\ensuremath{\Ppi}\xspace}

\def\pim   {\ensuremath{\pion^-}\xspace}

\def\kaon  {\ensuremath{\PK}\xspace}
  \def\Kbar  {\kern 0.2em\overline{\kern -0.2em \PK}{}\xspace}

\def\Kz    {\ensuremath{\kaon^0}\xspace}
\def\Kzb   {\ensuremath{\Kbar^0}\xspace}
\def\KzKzb {\ensuremath{\Kz \kern -0.16em \Kzb}\xspace}
\def\Kp    {\ensuremath{\kaon^+}\xspace}
\def\Km    {\ensuremath{\kaon^-}\xspace}

\def\KpKm  {\ensuremath{\Kp \kern -0.16em \Km}\xspace}

  \def\Dbar    {\kern 0.2em\overline{\kern -0.2em \PD}{}\xspace}
\def\D       {\ensuremath{\PD}\xspace}

\def\Dz      {\ensuremath{\D^0}\xspace}
\def\Dzb     {\ensuremath{\Dbar^0}\xspace}
\def\DzDzb   {\ensuremath{\Dz {\kern -0.16em \Dzb}}\xspace}
\def\Dp      {\ensuremath{\D^+}\xspace}
\def\Dm      {\ensuremath{\D^-}\xspace}

\def\DpDm    {\ensuremath{\Dp {\kern -0.16em \Dm}}\xspace}

\def\B       {\ensuremath{\PB}\xspace}
  \def\Bbar    {\kern 0.18em\overline{\kern -0.18em \PB}{}\xspace}

\def\Bu      {\ensuremath{\B^+}\xspace}

\def\Bd      {\ensuremath{\B^0}\xspace}
\def\Bs      {\ensuremath{\B^0_s}\xspace}

\def\jpsi     {\ensuremath{{\PJ\mskip -3mu/\mskip -2mu\Ppsi\mskip 2mu}}\xspace}
\def\psitwos  {\ensuremath{\Ppsi{(2S)}}\xspace}

  \def\Y#1S{\ensuremath{\PUpsilon{(#1S)}}\xspace}
\def\OneS  {\Y1S}
\def\TwoS  {\Y2S}
\def\ThreeS{\Y3S}

\newcommand{\tev}{\ensuremath{\mathrm{\,Te\kern -0.1em V}}\xspace}
\newcommand{\gev}{\ensuremath{\mathrm{\,Ge\kern -0.1em V}}\xspace}
\newcommand{\mev}{\ensuremath{\mathrm{\,Me\kern -0.1em V}}\xspace}
\newcommand{\kev}{\ensuremath{\mathrm{\,ke\kern -0.1em V}}\xspace}
\newcommand{\ev}{\ensuremath{\mathrm{\,e\kern -0.1em V}}\xspace}
\newcommand{\gevc}{\ensuremath{{\mathrm{\,Ge\kern -0.1em V\!/}c}}\xspace}
\newcommand{\mevc}{\ensuremath{{\mathrm{\,Me\kern -0.1em V\!/}c}}\xspace}
\newcommand{\gevcc}{\ensuremath{{\mathrm{\,Ge\kern -0.1em V\!/}c^2}}\xspace}
\newcommand{\gevgevcccc}{\ensuremath{{\mathrm{\,Ge\kern -0.1em V^2\!/}c^4}}\xspace}
\newcommand{\mevcc}{\ensuremath{{\mathrm{\,Me\kern -0.1em V\!/}c^2}}\xspace}

\def\mub{\ensuremath{\rm \,\upmu b}\xspace}

\def\invpb {\ensuremath{\mbox{\,pb}^{-1}}\xspace}

\def\invfb   {\ensuremath{\mbox{\,fb}^{-1}}\xspace}

\def\BR         {{\ensuremath{\cal B}\xspace}}

\newcommand{\decay}[2]{\ensuremath{#1\!\to #2}\xspace}         

\def\to                 {\ensuremath{\rightarrow}\xspace}

\def\gsim{{~\raise.15em\hbox{$>$}\kern-.85em
          \lower.35em\hbox{$\sim$}~}\xspace}
\def\lsim{{~\raise.15em\hbox{$<$}\kern-.85em
          \lower.35em\hbox{$\sim$}~}\xspace}

\def\BdToKpi      {\decay{\Bd}{\Kp\pim}\xspace}

\def\AT#1     {\ensuremath{A_T^{#1}}\xspace}           

\def\Bsmm     {\decay{\Bs}{\mup\mun}\xspace}
\def\Bdmm     {\decay{\Bd}{\mup\mun}\xspace}

\def\C#1      {\ensuremath{\mathcal{C}_{#1}}}                       
\def\Cp#1     {\ensuremath{\mathcal{C}_{#1}^{'}}}                    
\def\Ceff#1   {\ensuremath{\mathcal{C}_{#1}^{\mathrm{(eff)}}}}        
\def\Cpeff#1  {\ensuremath{\mathcal{C}_{#1}^{'\mathrm{(eff)}}}}       
\def\Ope#1    {\ensuremath{\mathcal{O}_{#1}}}                       
\def\Opep#1   {\ensuremath{\mathcal{O}_{#1}^{'}}}                    

\newcommand{\CL}{C.L.\ }

\newcommand{\CLs}{\ensuremath{\textrm{CL}_{\textrm{s}}}\xspace}
\newcommand{\CLb}{\ensuremath{\textrm{CL}_{\textrm{b}}}\xspace}

\newcommand{\bbdim}{\ensuremath{b\bar{b}\to \mu \mu X}\xspace}

\newcommand{\Bq}{\ensuremath{B^0_q}\xspace}
\newcommand{\Lambdab}{\ensuremath{\Lambda^0_b}\xspace}
\newcommand{\Bsmumu}{\ensuremath{\Bs\to\mu^+\mu^-}\xspace}
\newcommand{\Bdmumu}{\ensuremath{\Bd\to\mu^+\mu^-}\xspace}

\newcommand{\DKpi}{\ensuremath{\D^0\to K^-\pi^+}\xspace}

\newcommand{\Bdpipi}{\ensuremath{\Bd\to\pi^+\pi^-}\xspace}

\newcommand{\BsKK}{\ensuremath{\Bs\to K^+K^-}\xspace}

\newcommand{\BdKpi}{\ensuremath{\Bd\to K^+\pi^-}\xspace}

\newcommand{\Bhh}{\ensuremath{B^0_{q}\to h^+h^{'-}}\xspace}

\newcommand{\BuJpsiK}{\ensuremath{B^+\to J/\psi K^+}\xspace}
\newcommand{\BuJpsimumuK}{\ensuremath{B^+\to J/\psi(\mu^+\mu^-)K^+}\xspace}

\newcommand{\BdJpsiKst}{\ensuremath{B^0\to J/\psi K^{*0}}\xspace}
\newcommand{\BdJpsimumuKstKpi}{\ensuremath{B^0\to J/\psi(\mu^+\mu^-)K^{*0}(K^+\pi^-)}\xspace}

\newcommand{\Jpsi}{\ensuremath{J/\psi}\xspace}

\newcommand{\Bqmumu}{\ensuremath{\ensuremath{B^0_{q}}\to\mu^+\mu^-}\xspace}

\newcommand{\BsJpsimumuPhiKK}{\ensuremath{B^0_s\to J/\psi(\mu^+\mu^-) \phi(K^+K^-)}\xspace}

\newcommand{\BsJpsiPhi}{\ensuremath{B^0_s\to J/\psi \phi}\xspace}

\newcommand{\BRof}[1]{\ensuremath{{\cal B}(#1)}\xspace}

\newcommand{\MeVcc}{\ensuremath{\,{\rm MeV}/c^2}\xspace}
\newcommand{\IP}{\ensuremath{{\rm IP}}\xspace}

\newcommand{\gl}{\ensuremath{{\rm GL}}\xspace}

\newcommand\TTstrut{\rule{0pt}{3.2ex}}
\newcommand\Bstrut{\rule[-1.2ex]{0pt}{0pt}}
\newcommand\BBstrut{\rule[-1.8ex]{0pt}{0pt}}

\begin{document}

\begin{titlepage}

\belowpdfbookmark{Title page}{title}

\pagenumbering{roman}

\vspace*{-1.5cm}

\centerline{\large EUROPEAN ORGANIZATION FOR NUCLEAR RESEARCH (CERN)}

\vspace*{1.5cm}

\hspace*{-5mm}\begin{tabular*}{16cm}{lc@{\extracolsep{\fill}}r}

\vspace*{-12mm}\mbox{\!\!\!\epsfig{figure=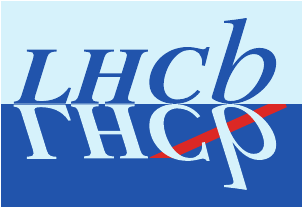,width=.12\textwidth}}& & \\

&& CERN-PH-EP-2011-029\\

&& 8 April 2011 \\

\end{tabular*}

\vspace*{4cm}

\begin{center}

{\bf\huge\boldmath {Search for the rare decays \Bsmumu and \Bdmumu}\\
}

\vspace*{2cm}

\normalsize {

The LHCb Collaboration\footnote{Authors are listed on the following pages.}


}

\end{center}

\vspace{\fill}

\centerline{\bf Abstract}

\vspace*{5mm}\noindent

A search for the decays \Bsmumu and \Bdmumu is performed with 
about 37 pb$^{-1}$ of $pp$ collisions at $\sqrt{s}$ = 7~TeV
collected by the LHCb experiment at the Large Hadron Collider at CERN.
The observed numbers of events are consistent with the background expectations. 
The resulting upper limits on the branching ratios are \BRof \Bsmumu $< 5.6 \times 10^{-8}$ 
and \BRof \Bdmumu $<1.5 \times 10^{-8}$ at 95\% confidence level.

\vspace*{1.cm}



\vspace{\fill}

\centerline{(Submitted to Physics Letters B.)}
\vspace*{0.5cm}

\end{titlepage}


\setcounter{page}{2}



\belowpdfbookmark{LHCb author list}{authors}

\centerline{\large\bf The LHCb Collaboration}
\begin{flushleft}
\small
R.~Aaij$^{23}$, 
B.~Adeva$^{36}$, 
M.~Adinolfi$^{42}$, 
C.~Adrover$^{6}$, 
A.~Affolder$^{48}$, 
Z.~Ajaltouni$^{5}$, 
J.~Albrecht$^{37}$, 
F.~Alessio$^{6,37}$, 
M.~Alexander$^{47}$, 
P.~Alvarez~Cartelle$^{36}$, 
A.A.~Alves~Jr$^{22}$, 
S.~Amato$^{2}$, 
Y.~Amhis$^{38}$, 
J.~Amoraal$^{23}$, 
J.~Anderson$^{39}$, 
R.B.~Appleby$^{50}$, 
O.~Aquines~Gutierrez$^{10}$, 
L.~Arrabito$^{53}$, 
M.~Artuso$^{52}$, 
E.~Aslanides$^{6}$, 
G.~Auriemma$^{22,m}$, 
S.~Bachmann$^{11}$, 
D.S.~Bailey$^{50}$, 
V.~Balagura$^{30,37}$, 
W.~Baldini$^{16}$, 
R.J.~Barlow$^{50}$, 
C.~Barschel$^{37}$, 
S.~Barsuk$^{7}$, 
A.~Bates$^{47}$, 
C.~Bauer$^{10}$, 
Th.~Bauer$^{23}$, 
A.~Bay$^{38}$, 
I.~Bediaga$^{1}$, 
K.~Belous$^{34}$, 
I.~Belyaev$^{30,37}$, 
E.~Ben-Haim$^{8}$, 
M.~Benayoun$^{8}$, 
G.~Bencivenni$^{18}$, 
R.~Bernet$^{39}$, 
M.-O.~Bettler$^{17,37}$, 
M.~van~Beuzekom$^{23}$, 
A.~Bien$^{11}$, 
S.~Bifani$^{12}$, 
A.~Bizzeti$^{17,h}$, 
P.M.~Bj\o rnstad$^{50}$, 
T.~Blake$^{49}$, 
F.~Blanc$^{38}$, 
C.~Blanks$^{49}$, 
J.~Blouw$^{11}$, 
S.~Blusk$^{52}$, 
A.~Bobrov$^{33}$, 
V.~Bocci$^{22}$, 
A.~Bondar$^{33}$, 
N.~Bondar$^{29}$, 
W.~Bonivento$^{15}$, 
S.~Borghi$^{47}$, 
A.~Borgia$^{52}$, 
E.~Bos$^{23}$, 
T.J.V.~Bowcock$^{48}$, 
C.~Bozzi$^{16}$, 
T.~Brambach$^{9}$, 
J.~van~den~Brand$^{24}$, 
J.~Bressieux$^{38}$, 
S.~Brisbane$^{51}$, 
M.~Britsch$^{10}$, 
T.~Britton$^{52}$, 
N.H.~Brook$^{42}$, 
H.~Brown$^{48}$, 
A.~B\"{u}chler-Germann$^{39}$, 
A.~Bursche$^{39}$, 
J.~Buytaert$^{37}$, 
S.~Cadeddu$^{15}$, 
J.M.~Caicedo~Carvajal$^{37}$, 
O.~Callot$^{7}$, 
M.~Calvi$^{20,j}$, 
M.~Calvo~Gomez$^{35,n}$, 
A.~Camboni$^{35}$, 
P.~Campana$^{18}$, 
A.~Carbone$^{14}$, 
G.~Carboni$^{21,k}$, 
R.~Cardinale$^{19,i}$, 
A.~Cardini$^{15}$, 
L.~Carson$^{36}$, 
K.~Carvalho~Akiba$^{23}$, 
G.~Casse$^{48}$, 
M.~Cattaneo$^{37}$, 
M.~Charles$^{51}$, 
Ph.~Charpentier$^{37}$, 
N.~Chiapolini$^{39}$, 
X.~Cid~Vidal$^{36}$, 
P.J.~Clark$^{46}$, 
P.E.L.~Clarke$^{46}$, 
M.~Clemencic$^{37}$, 
H.V.~Cliff$^{43}$, 
J.~Closier$^{37}$, 
C.~Coca$^{28}$, 
V.~Coco$^{23}$, 
J.~Cogan$^{6}$, 
P.~Collins$^{37}$, 
F.~Constantin$^{28}$, 
G.~Conti$^{38}$, 
A.~Contu$^{51}$, 
M.~Coombes$^{42}$, 
G.~Corti$^{37}$, 
G.A.~Cowan$^{38}$, 
R.~Currie$^{46}$, 
B.~D'Almagne$^{7}$, 
C.~D'Ambrosio$^{37}$, 
W.~Da~Silva$^{8}$, 
P.~David$^{8}$, 
I.~De~Bonis$^{4}$, 
S.~De~Capua$^{21,k}$, 
M.~De~Cian$^{39}$, 
F.~De~Lorenzi$^{12}$, 
J.M.~De~Miranda$^{1}$, 
L.~De~Paula$^{2}$, 
P.~De~Simone$^{18}$, 
D.~Decamp$^{4}$, 
H.~Degaudenzi$^{38,37}$, 
M.~Deissenroth$^{11}$, 
L.~Del~Buono$^{8}$, 
C.~Deplano$^{15}$, 
O.~Deschamps$^{5}$, 
F.~Dettori$^{15,d}$, 
J.~Dickens$^{43}$, 
H.~Dijkstra$^{37}$, 
M.~Dima$^{28}$, 
P.~Diniz~Batista$^{1}$, 
S.~Donleavy$^{48}$, 
P.~Dornan$^{49}$, 
D.~Dossett$^{44}$, 
A.~Dovbnya$^{40}$, 
F.~Dupertuis$^{38}$, 
R.~Dzhelyadin$^{34}$, 
C.~Eames$^{49}$, 
S.~Easo$^{45}$, 
U.~Egede$^{49}$, 
V.~Egorychev$^{30}$, 
S.~Eidelman$^{33}$, 
D.~van~Eijk$^{23}$, 
F.~Eisele$^{11}$, 
S.~Eisenhardt$^{46}$, 
L.~Eklund$^{47}$, 
Ch.~Elsasser$^{39}$, 
D.G.~d'Enterria$^{35,o}$, 
D.~Esperante~Pereira$^{36}$, 
L.~Est\`{e}ve$^{43}$, 
A.~Falabella$^{16,e}$, 
E.~Fanchini$^{20,j}$, 
C.~F\"{a}rber$^{11}$, 
G.~Fardell$^{46}$, 
C.~Farinelli$^{23}$, 
S.~Farry$^{12}$, 
V.~Fave$^{38}$, 
V.~Fernandez~Albor$^{36}$, 
M.~Ferro-Luzzi$^{37}$, 
S.~Filippov$^{32}$, 
C.~Fitzpatrick$^{46}$, 
F.~Fontanelli$^{19,i}$, 
R.~Forty$^{37}$, 
M.~Frank$^{37}$, 
C.~Frei$^{37}$, 
M.~Frosini$^{17,f,37}$, 
S.~Furcas$^{20}$, 
A.~Gallas~Torreira$^{36}$, 
D.~Galli$^{14,c}$, 
M.~Gandelman$^{2}$, 
P.~Gandini$^{51}$, 
Y.~Gao$^{3}$, 
J-C.~Garnier$^{37}$, 
J.~Garofoli$^{52}$, 
L.~Garrido$^{35}$, 
C.~Gaspar$^{37}$, 
N.~Gauvin$^{38}$, 
M.~Gersabeck$^{37}$, 
T.~Gershon$^{44}$, 
Ph.~Ghez$^{4}$, 
V.~Gibson$^{43}$, 
V.V.~Gligorov$^{37}$, 
C.~G\"{o}bel$^{54}$, 
D.~Golubkov$^{30}$, 
A.~Golutvin$^{49,30,37}$, 
A.~Gomes$^{2}$, 
H.~Gordon$^{51}$,
C.~Gotti$^{20}$, 
M.~Grabalosa~G\'{a}ndara$^{35}$, 
R.~Graciani~Diaz$^{35}$, 
L.A.~Granado~Cardoso$^{37}$, 
E.~Graug\'{e}s$^{35}$, 
G.~Graziani$^{17}$, 
A.~Grecu$^{28}$, 
S.~Gregson$^{43}$, 
B.~Gui$^{52}$, 
E.~Gushchin$^{32}$, 
Yu.~Guz$^{34}$, 
T.~Gys$^{37}$, 
G.~Haefeli$^{38}$, 
S.C.~Haines$^{43}$, 
T.~Hampson$^{42}$, 
S.~Hansmann-Menzemer$^{11}$, 
R.~Harji$^{49}$, 
N.~Harnew$^{51}$, 
P.F.~Harrison$^{44}$, 
J.~He$^{7}$, 
K.~Hennessy$^{48}$, 
P.~Henrard$^{5}$, 
J.A.~Hernando~Morata$^{36}$, 
E.~van~Herwijnen$^{37}$, 
A.~Hicheur$^{38}$, 
E.~Hicks$^{48}$, 
W.~Hofmann$^{10}$, 
K.~Holubyev$^{11}$, 
P.~Hopchev$^{4}$, 
W.~Hulsbergen$^{23}$, 
P.~Hunt$^{51}$, 
T.~Huse$^{48}$, 
R.S.~Huston$^{12}$, 
D.~Hutchcroft$^{48}$, 
V.~Iakovenko$^{7,41}$, 
P.~Ilten$^{12}$, 
J.~Imong$^{42}$, 
R.~Jacobsson$^{37}$, 
M.~Jahjah~Hussein$^{5}$, 
E.~Jans$^{23}$, 
F.~Jansen$^{23}$, 
P.~Jaton$^{38}$, 
B.~Jean-Marie$^{7}$, 
F.~Jing$^{3}$, 
M.~John$^{51}$, 
D.~Johnson$^{51}$, 
C.R.~Jones$^{43}$, 
B.~Jost$^{37}$, 
F.~Kapusta$^{8}$, 
T.M.~Karbach$^{9}$, 
J.~Keaveney$^{12}$, 
U.~Kerzel$^{37}$, 
T.~Ketel$^{24}$, 
A.~Keune$^{38}$, 
B.~Khanji$^{6}$, 
Y.M.~Kim$^{46}$, 
M.~Knecht$^{38}$, 
S.~Koblitz$^{37}$, 
A.~Konoplyannikov$^{30}$, 
P.~Koppenburg$^{23}$, 
A.~Kozlinskiy$^{23}$, 
L.~Kravchuk$^{32}$, 
G.~Krocker$^{11}$, 
P.~Krokovny$^{11}$, 
F.~Kruse$^{9}$, 
K.~Kruzelecki$^{37}$, 
M.~Kucharczyk$^{25}$, 
S.~Kukulak$^{25}$, 
R.~Kumar$^{14,37}$, 
T.~Kvaratskheliya$^{30,37}$, 
V.N.~La~Thi$^{38}$, 
D.~Lacarrere$^{37}$, 
G.~Lafferty$^{50}$, 
A.~Lai$^{15}$, 
R.W.~Lambert$^{37}$, 
G.~Lanfranchi$^{18}$, 
C.~Langenbruch$^{11}$, 
T.~Latham$^{44}$, 
R.~Le~Gac$^{6}$, 
J.~van~Leerdam$^{23}$, 
J.-P.~Lees$^{4}$, 
R.~Lef\`{e}vre$^{5}$, 
A.~Leflat$^{31,37}$, 
J.~Lefran\c{c}ois$^{7}$, 
O.~Leroy$^{6}$, 
T.~Lesiak$^{25}$, 
L.~Li$^{3}$, 
Y.Y.~Li$^{43}$, 
L.~Li~Gioi$^{5}$, 
M.~Lieng$^{9}$, 
M.~Liles$^{48}$, 
R.~Lindner$^{37}$, 
C.~Linn$^{11}$, 
B.~Liu$^{3}$, 
G.~Liu$^{37}$, 
J.H.~Lopes$^{2}$, 
E.~Lopez~Asamar$^{35}$, 
N.~Lopez-March$^{38}$, 
J.~Luisier$^{38}$, 
B.~M'charek$^{24}$, 
F.~Machefert$^{7}$, 
I.V.~Machikhiliyan$^{4,30}$, 
F.~Maciuc$^{10}$, 
O.~Maev$^{29,37}$, 
J.~Magnin$^{1}$, 
A.~Maier$^{37}$, 
S.~Malde$^{51}$, 
R.M.D.~Mamunur$^{37}$, 
G.~Manca$^{15,d,37}$, 
G.~Mancinelli$^{6}$, 
N.~Mangiafave$^{43}$, 
U.~Marconi$^{14}$, 
R.~M\"{a}rki$^{38}$, 
J.~Marks$^{11}$, 
G.~Martellotti$^{22}$, 
A.~Martens$^{7}$, 
L.~Martin$^{51}$, 
A.~Mart\'{i}n~S\'{a}nchez$^{7}$, 
D.~Martinez~Santos$^{37}$, 
A.~Massafferri$^{1}$, 
Z.~Mathe$^{12}$, 
C.~Matteuzzi$^{20}$, 
M.~Matveev$^{29}$, 
V.~Matveev$^{34}$, 
E.~Maurice$^{6}$, 
B.~Maynard$^{52}$, 
A.~Mazurov$^{32,37}$, 
G.~McGregor$^{50}$, 
R.~McNulty$^{12}$, 
C.~Mclean$^{46}$, 
M.~Meissner$^{11}$, 
M.~Merk$^{23}$, 
J.~Merkel$^{9}$, 
M.~Merkin$^{31}$, 
R.~Messi$^{21,k}$, 
S.~Miglioranzi$^{37}$, 
D.A.~Milanes$^{13,37}$, 
M.-N.~Minard$^{4}$, 
S.~Monteil$^{5}$, 
D.~Moran$^{12}$, 
P.~Morawski$^{25}$, 
J.V.~Morris$^{45}$, 
R.~Mountain$^{52}$, 
I.~Mous$^{23}$, 
F.~Muheim$^{46}$, 
K.~M\"{u}ller$^{39}$, 
R.~Muresan$^{28,38}$, 
F.~Murtas$^{18}$, 
B.~Muryn$^{26}$, 
M.~Musy$^{35}$, 
J.~Mylroie-Smith$^{48}$, 
P.~Naik$^{42}$, 
T.~Nakada$^{38}$, 
R.~Nandakumar$^{45}$, 
J.~Nardulli$^{45}$, 
M.~Nedos$^{9}$, 
M.~Needham$^{46}$, 
N.~Neufeld$^{37}$, 
M.~Nicol$^{7}$, 
S.~Nies$^{9}$, 
V.~Niess$^{5}$, 
N.~Nikitin$^{31}$, 
A.~Oblakowska-Mucha$^{26}$, 
V.~Obraztsov$^{34}$, 
S.~Oggero$^{23}$, 
O.~Okhrimenko$^{41}$, 
R.~Oldeman$^{15,d}$, 
M.~Orlandea$^{28}$, 
A.~Ostankov$^{34}$, 
B.~Pal$^{52}$, 
J.~Palacios$^{39}$, 
M.~Palutan$^{18}$, 
J.~Panman$^{37}$, 
A.~Papanestis$^{45}$, 
M.~Pappagallo$^{13,b}$, 
C.~Parkes$^{47,37}$, 
C.J.~Parkinson$^{49}$, 
G.~Passaleva$^{17}$, 
G.D.~Patel$^{48}$, 
M.~Patel$^{49}$, 
S.K.~Paterson$^{49,37}$, 
G.N.~Patrick$^{45}$, 
C.~Patrignani$^{19,i}$, 
C.~Pavel-Nicorescu$^{28}$, 
A.~Pazos~Alvarez$^{36}$, 
A.~Pellegrino$^{23}$, 
G.~Penso$^{22,l}$, 
M.~Pepe~Altarelli$^{37}$, 
S.~Perazzini$^{14,c}$, 
D.L.~Perego$^{20,j}$, 
E.~Perez~Trigo$^{36}$, 
A.~P\'{e}rez-Calero~Yzquierdo$^{35}$, 
P.~Perret$^{5}$, 
A.~Petrella$^{16,e,37}$, 
A.~Petrolini$^{19,i}$, 
B.~Pie~Valls$^{35}$, 
B.~Pietrzyk$^{4}$, 
D.~Pinci$^{22}$, 
R.~Plackett$^{47}$, 
S.~Playfer$^{46}$, 
M.~Plo~Casasus$^{36}$, 
G.~Polok$^{25}$, 
A.~Poluektov$^{44,33}$, 
E.~Polycarpo$^{2}$, 
D.~Popov$^{10}$, 
B.~Popovici$^{28}$, 
C.~Potterat$^{38}$, 
A.~Powell$^{51}$, 
T.~du~Pree$^{23}$, 
V.~Pugatch$^{41}$, 
A.~Puig~Navarro$^{35}$, 
W.~Qian$^{3}$, 
J.H.~Rademacker$^{42}$, 
B.~Rakotomiaramanana$^{38}$, 
I.~Raniuk$^{40}$, 
G.~Raven$^{24}$, 
S.~Redford$^{51}$, 
W.~Reece$^{49}$, 
A.C.~dos~Reis$^{1}$, 
S.~Ricciardi$^{45}$, 
K.~Rinnert$^{48}$, 
D.A.~Roa~Romero$^{5}$, 
P.~Robbe$^{7}$, 
E.~Rodrigues$^{47}$, 
F.~Rodrigues$^{2}$, 
C.~Rodriguez~Cobo$^{36}$, 
P.~Rodriguez~Perez$^{36}$, 
G.J.~Rogers$^{43}$, 
V.~Romanovsky$^{34}$, 
J.~Rouvinet$^{38}$, 
T.~Ruf$^{37}$, 
H.~Ruiz$^{35}$, 
G.~Sabatino$^{21,k}$, 
J.J.~Saborido~Silva$^{36}$, 
N.~Sagidova$^{29}$, 
P.~Sail$^{47}$, 
B.~Saitta$^{15,d}$, 
C.~Salzmann$^{39}$, 
A.~Sambade~Varela$^{37}$, 
M.~Sannino$^{19,i}$, 
R.~Santacesaria$^{22}$, 
R.~Santinelli$^{37}$, 
E.~Santovetti$^{21,k}$, 
M.~Sapunov$^{6}$, 
A.~Sarti$^{18}$, 
C.~Satriano$^{22,m}$, 
A.~Satta$^{21}$, 
M.~Savrie$^{16,e}$, 
D.~Savrina$^{30}$, 
P.~Schaack$^{49}$, 
M.~Schiller$^{11}$, 
S.~Schleich$^{9}$, 
M.~Schmelling$^{10}$, 
B.~Schmidt$^{37}$, 
O.~Schneider$^{38}$, 
A.~Schopper$^{37}$, 
M.-H.~Schune$^{7}$, 
R.~Schwemmer$^{37}$, 
A.~Sciubba$^{18,l}$, 
M.~Seco$^{36}$, 
A.~Semennikov$^{30}$, 
K.~Senderowska$^{26}$, 
N.~Serra$^{23}$, 
J.~Serrano$^{6}$, 
B.~Shao$^{3}$, 
M.~Shapkin$^{34}$, 
I.~Shapoval$^{40,37}$, 
P.~Shatalov$^{30}$, 
Y.~Shcheglov$^{29}$, 
T.~Shears$^{48}$, 
L.~Shekhtman$^{33}$, 
O.~Shevchenko$^{40}$, 
V.~Shevchenko$^{30}$, 
A.~Shires$^{49}$, 
E.~Simioni$^{24}$, 
H.P.~Skottowe$^{43}$, 
T.~Skwarnicki$^{52}$, 
A.C.~Smith$^{37}$, 
K.~Sobczak$^{5}$, 
F.J.P.~Soler$^{47}$, 
A.~Solomin$^{42}$, 
P.~Somogy$^{37}$, 
F.~Soomro$^{49}$, 
B.~Souza~De~Paula$^{2}$, 
B.~Spaan$^{9}$, 
A.~Sparkes$^{46}$, 
E.~Spiridenkov$^{29}$, 
P.~Spradlin$^{51}$, 
F.~Stagni$^{37}$, 
O.~Steinkamp$^{39}$, 
O.~Stenyakin$^{34}$, 
S.~Stoica$^{28}$, 
S.~Stone$^{52}$, 
B.~Storaci$^{23}$, 
U.~Straumann$^{39}$, 
N.~Styles$^{46}$, 
M.~Szczekowski$^{27}$, 
P.~Szczypka$^{38}$, 
T.~Szumlak$^{26}$, 
S.~T'Jampens$^{4}$, 
V.~Talanov$^{34}$, 
E.~Teodorescu$^{28}$, 
F.~Teubert$^{37}$, 
C.~Thomas$^{51,45}$, 
E.~Thomas$^{37}$, 
J.~van~Tilburg$^{39}$, 
V.~Tisserand$^{4}$, 
M.~Tobin$^{39}$, 
S.~Topp-Joergensen$^{51}$, 
M.T.~Tran$^{38}$, 
A.~Tsaregorodtsev$^{6}$, 
N.~Tuning$^{23}$, 
A.~Ukleja$^{27}$, 
P.~Urquijo$^{52}$, 
U.~Uwer$^{11}$, 
V.~Vagnoni$^{14}$, 
G.~Valenti$^{14}$, 
R.~Vazquez~Gomez$^{35}$, 
P.~Vazquez~Regueiro$^{36}$, 
S.~Vecchi$^{16}$, 
J.J.~Velthuis$^{42}$, 
M.~Veltri$^{17,g}$, 
K.~Vervink$^{37}$, 
B.~Viaud$^{7}$, 
I.~Videau$^{7}$, 
X.~Vilasis-Cardona$^{35,n}$, 
J.~Visniakov$^{36}$, 
A.~Vollhardt$^{39}$, 
D.~Voong$^{42}$, 
A.~Vorobyev$^{29}$, 
An.~Vorobyev$^{29}$, 
H.~Voss$^{10}$, 
K.~Wacker$^{9}$, 
S.~Wandernoth$^{11}$, 
J.~Wang$^{52}$, 
D.R.~Ward$^{43}$, 
A.D.~Webber$^{50}$, 
M.~Whitehead$^{44}$, 
D.~Wiedner$^{11}$, 
L.~Wiggers$^{23}$, 
G.~Wilkinson$^{51}$, 
M.P.~Williams$^{44,45}$, 
M.~Williams$^{49}$, 
F.F.~Wilson$^{45}$, 
J.~Wishahi$^{9}$, 
M.~Witek$^{25}$, 
W.~Witzeling$^{37}$, 
S.A.~Wotton$^{43}$, 
K.~Wyllie$^{37}$, 
Y.~Xie$^{46}$, 
F.~Xing$^{51}$, 
Z.~Yang$^{3}$, 
G.~Ybeles~Smit$^{23}$, 
R.~Young$^{46}$, 
O.~Yushchenko$^{34}$, 
M.~Zavertyaev$^{10,a}$, 
L.~Zhang$^{52}$, 
W.C.~Zhang$^{12}$, 
Y.~Zhang$^{3}$, 
A.~Zhelezov$^{11}$, 
L.~Zhong$^{3}$, 
E.~Zverev$^{31}$.\bigskip\newline{\it
\footnotesize
$ ^{1}$Centro Brasileiro de Pesquisas F\'{i}sicas (CBPF), Rio de Janeiro, Brazil\\
$ ^{2}$Universidade Federal do Rio de Janeiro (UFRJ), Rio de Janeiro, Brazil\\
$ ^{3}$Center for High Energy Physics, Tsinghua University, Beijing, China\\
$ ^{4}$LAPP, Universit\'{e} de Savoie, CNRS/IN2P3, Annecy-Le-Vieux, France\\
$ ^{5}$Clermont Universit\'{e}, Universit\'{e} Blaise Pascal, CNRS/IN2P3, LPC, Clermont-Ferrand, France\\
$ ^{6}$CPPM, Aix-Marseille Universit\'{e}, CNRS/IN2P3, Marseille, France\\
$ ^{7}$LAL, Universit\'{e} Paris-Sud, CNRS/IN2P3, Orsay, France\\
$ ^{8}$LPNHE, Universit\'{e} Pierre et Marie Curie, Universit\'{e} Paris Diderot, CNRS/IN2P3, Paris, France\\
$ ^{9}$Fakult\"{a}t Physik, Technische Universit\"{a}t Dortmund, Dortmund, Germany\\
$ ^{10}$Max-Planck-Institut f\"{u}r Kernphysik (MPIK), Heidelberg, Germany\\
$ ^{11}$Physikalisches Institut, Ruprecht-Karls-Universit\"{a}t Heidelberg, Heidelberg, Germany\\
$ ^{12}$School of Physics, University College Dublin, Dublin, Ireland\\
$ ^{13}$Sezione INFN di Bari, Bari, Italy\\
$ ^{14}$Sezione INFN di Bologna, Bologna, Italy\\
$ ^{15}$Sezione INFN di Cagliari, Cagliari, Italy\\
$ ^{16}$Sezione INFN di Ferrara, Ferrara, Italy\\
$ ^{17}$Sezione INFN di Firenze, Firenze, Italy\\
$ ^{18}$Laboratori Nazionali dell'INFN di Frascati, Frascati, Italy\\
$ ^{19}$Sezione INFN di Genova, Genova, Italy\\
$ ^{20}$Sezione INFN di Milano Bicocca, Milano, Italy\\
$ ^{21}$Sezione INFN di Roma Tor Vergata, Roma, Italy\\
$ ^{22}$Sezione INFN di Roma Sapienza, Roma, Italy\\
$ ^{23}$Nikhef National Institute for Subatomic Physics, Amsterdam, Netherlands\\
$ ^{24}$Nikhef National Institute for Subatomic Physics and Vrije Universiteit, Amsterdam, Netherlands\\
$ ^{25}$Henryk Niewodniczanski Institute of Nuclear Physics  Polish Academy of Sciences, Cracow, Poland\\
$ ^{26}$Faculty of Physics \& Applied Computer Science, Cracow, Poland\\
$ ^{27}$Soltan Institute for Nuclear Studies, Warsaw, Poland\\
$ ^{28}$Horia Hulubei National Institute of Physics and Nuclear Engineering, Bucharest-Magurele, Romania\\
$ ^{29}$Petersburg Nuclear Physics Institute (PNPI), Gatchina, Russia\\
$ ^{30}$Institute of Theoretical and Experimental Physics (ITEP), Moscow, Russia\\
$ ^{31}$Institute of Nuclear Physics, Moscow State University (SINP MSU), Moscow, Russia\\
$ ^{32}$Institute for Nuclear Research of the Russian Academy of Sciences (INR RAN), Moscow, Russia\\
$ ^{33}$Budker Institute of Nuclear Physics (BINP), Novosibirsk, Russia\\
$ ^{34}$Institute for High Energy Physics(IHEP), Protvino, Russia\\
$ ^{35}$Universitat de Barcelona, Barcelona, Spain\\
$ ^{36}$Universidad de Santiago de Compostela, Santiago de Compostela, Spain\\
$ ^{37}$European Organization for Nuclear Research (CERN), Geneva, Switzerland\\
$ ^{38}$Ecole Polytechnique F\'{e}d\'{e}rale de Lausanne (EPFL), Lausanne, Switzerland\\
$ ^{39}$Physik-Institut, Universit\"{a}t Z\"{u}rich, Z\"{u}rich, Switzerland\\
$ ^{40}$NSC Kharkiv Institute of Physics and Technology (NSC KIPT), Kharkiv, Ukraine\\
$ ^{41}$Institute for Nuclear Research of the National Academy of Sciences (KINR), Kyiv, Ukraine\\
$ ^{42}$H.H. Wills Physics Laboratory, University of Bristol, Bristol, United Kingdom\\
$ ^{43}$Cavendish Laboratory, University of Cambridge, Cambridge, United Kingdom\\
$ ^{44}$Department of Physics, University of Warwick, Coventry, United Kingdom\\
$ ^{45}$STFC Rutherford Appleton Laboratory, Didcot, United Kingdom\\
$ ^{46}$School of Physics and Astronomy, University of Edinburgh, Edinburgh, United Kingdom\\
$ ^{47}$School of Physics and Astronomy, University of Glasgow, Glasgow, United Kingdom\\
$ ^{48}$Oliver Lodge Laboratory, University of Liverpool, Liverpool, United Kingdom\\
$ ^{49}$Imperial College London, London, United Kingdom\\
$ ^{50}$School of Physics and Astronomy, University of Manchester, Manchester, United Kingdom\\
$ ^{51}$Department of Physics, University of Oxford, Oxford, United Kingdom\\
$ ^{52}$Syracuse University, Syracuse, NY, United States of America\\
$ ^{53}$CC-IN2P3, CNRS/IN2P3, Lyon-Villeurbanne, France, associated member\\
$ ^{54}$Pontif\'{i}cia Universidade Cat\'{o}lica do Rio de Janeiro (PUC-Rio), Rio de Janeiro, Brazil, associated to $^2 $\\
\bigskip
$ ^{a}$P.N. Lebedev Physical Institute, Russian Academy of Science (LPI RAS), Moskow, Russia\\
$ ^{b}$Universit\`{a} di Bari, Bari, Italy\\
$ ^{c}$Universit\`{a} di Bologna, Bologna, Italy\\
$ ^{d}$Universit\`{a} di Cagliari, Cagliari, Italy\\
$ ^{e}$Universit\`{a} di Ferrara, Ferrara, Italy\\
$ ^{f}$Universit\`{a} di Firenze, Firenze, Italy\\
$ ^{g}$Universit\`{a} di Urbino, Urbino, Italy\\
$ ^{h}$Universit\`{a} di Modena e Reggio Emilia, Modena, Italy\\
$ ^{i}$Universit\`{a} di Genova, Genova, Italy\\
$ ^{j}$Universit\`{a} di Milano Bicocca, Milano, Italy\\
$ ^{k}$Universit\`{a} di Roma Tor Vergata, Roma, Italy\\
$ ^{l}$Universit\`{a} di Roma La Sapienza, Roma, Italy\\
$ ^{m}$Universit\`{a} della Basilicata, Potenza, Italy\\
$ ^{n}$LIFAELS, La Salle, Universitat Ramon Llull, Barcelona, Spain\\
$ ^{o}$Instituci\'{o} Catalana de Recerca i Estudis Avan\c{c}ats (ICREA), Barcelona, Spain\\
}
\end{flushleft}

\cleardoublepage

\setcounter{page}{1}

\pagenumbering{arabic}

%


\pagestyle{plain} 
\setcounter{page}{1}
\pagenumbering{arabic}


%


\section{Introduction}
\label{sec:introduction}

Within the Standard Model (SM) exclusive dimuon decays of the \Bd and \Bs 
mesons\footnote{In this Letter the inclusion of charge-conjugate states is implicit.} are rare as they
occur only via loop diagrams and are helicity suppressed. 
New Physics models, especially those with an extended Higgs sector,
can significantly enhance the branching fractions, although in some models the rates are lowered.

The amplitudes contributing to the branching ratio 
\BRof\Bqmumu (where $q = d, s$ for the \Bd and \Bs mesons respectively) 
can be expressed in terms of the scalar ($c_S$), pseudoscalar ($c_P$) and axial vector ($c_A$) Wilson coefficients in a completely general approach~\cite{Bobeth2001}. 
Within the SM, the contributions of $c_{S}$ and $c_{P}$ are negligible while $c_A$ is 
calculated with an accuracy of a few percent~\cite{SMprediction}. 
The dominant contribution stems from an electroweak penguin with a \Z decaying into two muons.  
However, the accuracy of the SM prediction for \BRof \Bqmumu is limited by 
the knowledge of the decay constants of the $B^0_q$ mesons. This limitation can be reduced  
by normalizing to the well-measured mass differences of the $B^0_q$ mesons. 
Using this approach~\cite{Buras2010}, the SM predictions are 
\begin{eqnarray}
\label{BR:bqll_SMpred}
\BRof\Bsmumu_{\rm SM}  & = &  (0.32 \pm 0.02) \times 10^{-8} \ , \nonumber \\
\BRof\Bdmumu_{\rm SM}  & = &  (0.010 \pm 0.001) \times 10^{-8} \nonumber \ .
\end{eqnarray}
Many extensions to the SM predict a very different Higgs sector. For instance, within the Minimal Supersymmetric SM 
(MSSM) in the large $\tan \beta$ approximation~\cite{MSSM}, $c_{S,P}^{\rm MSSM} \propto \tan^3\beta/M_A^2$, 
where $M_A$ denotes the pseudoscalar Higgs mass and $\tan \beta$ the ratio of Higgs vacuum expectation values.  
The most restrictive limits on the search for \Bqmumu
have so far been achieved at the Tevatron, due to the large $b \bar{b}$ 
cross-section at hadron colliders. 
The best limits at $95\%$ \CL published 
so far are obtained using $6.1~\invfb$ by the D0 collaboration~\cite{d0_PLB}, 
\BRof\Bsmumu $<5.1 \times 10^{-8}$, 
and using $2~\invfb$ by the CDF collaboration~\cite{cdf_PRL}, 
\BRof\Bsmumu $<5.8 \times 10^{-8}$ and 
\BRof\Bdmumu $<1.8 \times 10^{-8}$. 
The CDF collaboration has also presented 
preliminary results~\cite{cdf_preliminary} with $3.7~\invfb$, that lower the limits to \BRof\Bsmumu $<4.3 \times 10^{-8}$ and \BRof\Bdmumu $<0.76 \times 10^{-8}$.

The \lhcb experiment is well suited 
for such searches due to its good invariant mass resolution, 
vertex resolution, muon identification and trigger acceptance. 
In addition, \lhcb has a hadronic trigger capability which
provides large samples of \Bhh decays, where $h$ and $h'$ stand for a hadron (kaon or pion). 
These are used as control samples in order to reduce
the dependence of the results on 
the simulation. 

The measurements in this Letter use about 37~\invpb of integrated luminosity 
collected by LHCb between July and October 2010 at $\sqrt{s}$ = 7~TeV. 
Assuming the SM branching ratio, about $0.7~(0.08)$ \Bsmumu(\Bdmumu) are expected to be
reconstructed using the $b \bar{b}$ cross-section, measured
within the LHCb acceptance, of $75\pm14\,\mub$~\cite{bbxsection}. 

\section{The LHCb detector}
\label{sec:detector}

The LHCb detector~\cite{LHCbdetector}
is a single-arm forward spectrometer 
with an angular coverage 
from approximately 10 mrad to 300 (250) mrad in the bending (non-bending) plane.
The detector consists of a vertex locator (VELO), 
a warm dipole magnet with a bending power of $\int B dl =  4$ T$\,$m, a tracking system, two ring-imaging Cherenkov detectors (RICH),
a calorimeter system and a muon system. The VELO 
consists of a series of silicon modules, each providing a measure of the radial and azimuthal coordinates, with
the sensitive area starting at 8 mm from 
the beam line during collisions.
The tracking system comprises four layers of silicon sensors 
before the magnet and 
three stations equipped with silicon sensors
in the inner part 
and straw tubes in the outer part 
after the magnet. 
Track momenta are measured with a precision between
$\delta p / p = 0.35\%$ at 5~\gevc and $\delta p / p = 0.5\%$ at 100~\gevc. 
The RICH system 
provides charged hadron identification in a momentum range 2--100 GeV$/c$.
The calorimeter system consists of a preshower, a scintillating pad detector, 
an electromagnetic calorimeter 
and a hadronic 
calorimeter. It identifies high transverse energy ($E_{\rm T}$) hadron, 
electron and photon candidates and provides information for the trigger.
Five muon stations composed of MWPC (except in the highest rate region, where triple-GEMs are used) 
provide fast information
for the trigger and muon identification capability.

LHCb has a two-level trigger system both for leptonic and purely hadronic 
final states. It exploits the finite lifetime and relatively large mass of charm and beauty hadrons
to distinguish heavy flavour decays from the dominant light quark processes.
The first trigger level (L0) is implemented in hardware and reduces the rate
to a maximum of 1~MHz, the read-out rate of the whole detector.
The second trigger level (High Level Trigger, HLT) 
is implemented in software running on an event filter CPU farm. In the first stage of the 
software trigger (HLT1) a partial event reconstruction is performed. 
The second stage (HLT2) performs a full event reconstruction to enhance the signal purity further.

The forward geometry of LHCb allows the first level trigger to collect events containing one or two muons with 
very low transverse momenta ($p_{\rm T}$): more than 90\% of the data were collected with a $p_{\rm T}$ threshold of 1.4 GeV/$c$
for single muon triggers and $p_{\rm T}(\mu_1)>0.48$ GeV/$c$ and  $p_{\rm T}(\mu_2) > 0.56$ GeV/$c$ for dimuon triggers. 
The $E_{\rm T}$ threshold for the hadron trigger varied in the range 2.6 to 3.6~GeV.
The single muon trigger line in the HLT requires either 
$p_{\rm T}>1.8$ GeV/$c$ or includes a cut on the impact parameter (\IP) with respect to the 
primary vertex, which allows for a lower $p_{\rm T}$ requirement ($p_{\rm T}>0.8$ GeV/$c$, $\rm IP> 0.11$~mm). 
The dimuon trigger line requires muon pairs of opposite charge forming a common vertex 
and an invariant mass $M_{\mu\mu} > 4.7 $ GeV/$c^2$.   
A second trigger line, primarily to select \jpsi events, requires 
$2.97 < M_{\mu\mu} < 3.21$ GeV/$c^2$. 
The remaining region of the dimuon invariant mass is also covered by trigger lines that 
in addition require the dimuon secondary vertex to be well separated from the primary vertex.
Other HLT trigger lines select generic displaced vertices, providing a high 
efficiency for purely hadronic decays (for instance \Bhh).

\section{Analysis strategy}
\label{sec:strategy}

An important feature of this analysis is 
to rely as much as possible on data and to restrict to a minimum the use of simulation. 
Nevertheless, some Monte Carlo (MC) simulation has
been used, based on the PYTHIA 6.4 generator~\cite{PYTHIA} and the 
GEANT4 package~\cite{GEANT} for detector simulation.
The first part of the analysis is the event selection (Section~\ref{sec:selection}), 
which significantly reduces the size of the dataset by 
rejecting most of the background. 

The second part consists of the study of three 
normalization channels with known branching ratios: \BuJpsimumuK, \BsJpsimumuPhiKK and $\BdToKpi.$ 
Using each of these normalization channels, \BRof\Bqmumu  can be calculated as:
\begin{eqnarray}
\BRof\Bqmumu &=& {\BR_{\rm norm}}\times\frac{\rm
\epsilon_{norm}^{REC}
\epsilon_{norm}^{SEL|REC}
\epsilon_{norm}^{TRIG|SEL}
}{\rm
\epsilon_{sig}^{REC}
\epsilon_{sig}^{SEL|REC}
\epsilon_{sig}^{TRIG|SEL}
}\times\frac{f_{\rm norm}}{f_{\Bq}}
\times\frac{N_{\Bqmumu}}{N_{\rm norm}} \nonumber\\
&=& \alpha_{\Bqmumu} \times N_{\Bqmumu}\,,
\label{eq:normAlpha}
\end{eqnarray}
where $\alpha_{\Bqmumu}$ denotes the normalization factor,
$f_{\B^0_q}$  denotes the probability that a $b$-quark fragments 
into a \Bq and $f_{\rm norm}$ denotes the probability that a $b$-quark fragments 
into the $b$-hadron relevant for the chosen normalization channel with branching fraction 
\BR$_{\rm norm}$. The reconstruction 
efficiency ($\epsilon^{\rm REC}$) includes the acceptance and particle identification, while 
$\epsilon^{\rm SEL|REC}$ denotes the selection efficiency on reconstructed events.
The trigger efficiency on selected events is denoted by $\epsilon^{\rm TRIG|SEL}$.
This normalization ensures that knowledge of the absolute luminosity and $b \bar{b}$ 
production cross-section are not needed,
and that many systematic uncertainties cancel in the ratio of the efficiencies. The event selection 
for these channels is specifically designed to be as close as possible to the signal selection.
The ratios of reconstruction and selection efficiencies 
are estimated from the simulation, while 
the ratios of trigger 
efficiencies on selected events are determined from data (Section~\ref{sec:normalization}). 

In the third part of the analysis (Section~\ref{sec:likelihood}) 
each selected event is given a probability to be signal or background in a 
two-dimensional probability space defined by the dimuon invariant mass and 
a geometrical likelihood (\gl). 
The dimuon invariant mass and \gl probability density functions for both signal and background are 
determined from data.
This procedure ensures that even though the \gl is defined using simulated events, 
the result will not be biased by discrepancies between data and simulation.

Section~\ref{sec:results} describes the final measurement. 
In order to avoid unconscious bias in the analysis, the invariant mass region for the signal 
($M_{\Bd} \pm 60 \mevcc$ and $M_{\Bs} \pm 60 \mevcc$) was blinded until the selection criteria 
and analysis procedure had been defined.

\section{Event selection}
\label{sec:selection}

The selection has been designed in order to reduce the data sample to a manageable level
by simultaneously keeping the efficiency for the signals as high as possible and 
the selection between signals and control channels as similar as possible. 
This last requirement is needed to minimize the systematic uncertainty in the 
ratio of the selection efficiencies. The optimal separation between signal and background is left
to the likelihoods (Sect.~\ref{sec:likelihood}).
The basic cuts of the selection have been defined on Monte Carlo simulation \cite{roadmap} and then 
adapted to the data.

The data for the signal and all the normalization candidates 
are selected using either an inclusive two-body or a \Jpsi  
selection. 
Tracks are first 
required to be of good quality ($\chi^2/{\rm ndf}<5$) and to be displaced with respect to
the closest primary vertex (${\rm \chi_{IP}^2}/{\rm ndf}>12.5$, where $\rm\chi_{IP}^2$ is the difference
between the $\chi^2$ of the primary vertex built with and without the considered track).
To reject bad combinations before performing the vertex fit, the two tracks are
required to have a distance of closest approach of less than
0.3\,mm. The 
secondary vertex is required to be well fitted ($\chi^2/{\rm ndf}<9$)
and must be clearly separated from the primary in the forward
direction (vertex distance significance larger than 15). When more than one primary vertex is
reconstructed, the one that gives the minimum impact
parameter significance for the candidate is chosen.  The reconstructed candidate has to point to
the primary vertex ($\rm \chi_{IP}^2/{\rm ndf}<12.5$) in the case of the inclusive two-body
selection.  For all selections, the primary vertex is refitted excluding the signal
tracks before calculating the $\rm\chi_{IP}^2$/ndf and the vertex distance significance of the candidate. 

Tracks are defined as muons if they have at least one hit in two to four of the last four 
muon stations depending on the 
momentum. In the inclusive \Jpsi selection both tracks must be identified as muons 
and have an invariant mass 
within 60\,\MeVcc of the nominal \Jpsi mass.
The efficiency of the muon identification requirement has been measured using an inclusive
sample of \Jpsi events where one of the tracks does not use any information from
the muon chambers. The efficiency measured with data agrees with MC expectations
as a function of momentum within $2\%$,
and the residual differences are taken into account in the systematic uncertainties.

Events passing the two-body selection are considered 
\Bqmumu candidates if both tracks pass the muon identification 
criteria, 
and their invariant mass lies 
within 60\,\MeVcc of the nominal $B^0_q$ mass. The invariant mass of the \Bhh candidates 
has to be within 600\,\MeVcc of the nominal $B^0_q$ mass. 
As the acceptance of the tracking stations is larger than the muon chambers, the selected 
\Bhh candidates are required to have both tracks within the muon chamber acceptance to minimize the 
differences with \Bqmumu.  
The total efficiencies including acceptance, reconstruction 
and selection criteria on MC \Bqmumu and \Bhh events 
are $5.5\%$ and $4.5\%$ respectively; the main difference is due to material interactions. 
Assuming the SM branching ratio, 
0.3 \Bsmumu and 0.04 \Bdmumu events are expected after all selection requirements. There
are 343 (342) \Bqmumu candidates selected from data in the \Bs(\Bd) mass window.

The dominant background after the \Bqmumu selection is expected to be \bbdim
~\cite{roadmap, thesis2010_068}. This is confirmed by comparing the kinematical distributions of 
the sideband data with a \bbdim MC sample.
The muon misidentification probability as a function
of momentum obtained from data using $K^0_S \to \pi^+ \pi^-$, $\Lambda \to p \pi^{-} $ and $\phi
\to K^+K^-$ decays is in good agreement with MC expectations.~An estimate of the background
coming from misidentified hadrons is obtained by reweighting the hadron
misidentification probability using the momentum spectrum of the background in
the invariant mass sidebands.  The single hadron average misidentification probability is measured
to be $(7.1 \pm 0.5) \times 10^{-3}$ and the double hadron misidentification
probability is $(3.5\pm 0.9)\times 10^{-5}$, where the correlation between the momenta 
of the two hadrons is taken into account. 
About 10\% of the background is due to
pairs consisting of one real muon and a hadron misidentified as muon, mostly from decays in flight. 
The contribution from double misidentified hadrons is negligible. 
The number of expected \Bhh candidates
misidentified as \Bqmumu within the search window of $\pm 60 \mevcc$ around the
\Bs(\Bd) mass is less than 0.1 (0.3).

For the \BuJpsiK and \BsJpsiPhi normalization channels some additional cuts are required. 
In the former case, the $K^{\pm}$  candidates are required to pass the same track quality
and impact parameter cuts as the
muons from the \Jpsi. 
For \BsJpsiPhi candidates, the $K^+K^-$ invariant mass is required to be within $\pm 10 \mevcc$
of the $\phi$ mass~\cite{PDG}. The $B$ vertex has to be of good quality, $\chi^2/{\rm ndf} < 25$.
The requirements on $\rm\chi_{IP}^2$/ndf and 
vertex separation significance for the $B$ candidate are the same as those for the signal
selection. The total efficiencies including acceptance, reconstruction 
and selection criteria  
for MC \BuJpsiK  and \BsJpsiPhi events are $2.6\%$ and $1.3\%$ 
respectively.

\section{Evaluation of the normalization factor}
\label{sec:normalization}

The branching fractions of the three normalization channels, 
\BuJpsimumuK, 
\BsJpsimumuPhiKK and $\BdToKpi,$ are shown in Table~\ref{tab:norm_summary}.
The first two decays have similar trigger and muon identification efficiency to the signal 
but a different number of particles in the final state, while the third channel has 
the same two-body topology but is selected with the 
hadronic trigger. The branching ratio of the \BsJpsiPhi decay is not known precisely ($\sim 25\%$) 
but has the advantage that the normalization of \Bsmumu with a \Bs decay 
does not require the knowledge of the ratio of fragmentation fractions, which has an uncertainty of
$\sim 13\%$~\cite{HFAG}. 

\subsection{Ratio of reconstruction and selection efficiencies}
\label{sec:ratio_recsel}

The accuracy of the simulation of the reconstruction efficiency $\epsilon^{\rm REC}$ relies on 
the knowledge of the detector geometrical acceptance, the material interactions and the tracking efficiency. 
The uncertainty on the tracking efficiency is taken to be $4\%$ per track~\cite{bbxsection} and this
is the dominant source of the systematic uncertainty in the ratio with the two normalization 
channels involving \jpsi mesons. 
The ratios $\epsilon_{\rm norm}^{\rm REC}/\epsilon_{\rm sig}^{\rm REC}$ predicted by the simulation are
$0.58 \pm 0.02~(\BuJpsiK),$ $0.39 \pm 0.03~(\BsJpsiPhi)$ and $0.75 \pm 0.05~(\BdKpi)$. 

The effect of an extra particle on the ratio of $\epsilon^{\rm REC}$ is cross-checked
in data using the decay \BdJpsimumuKstKpi.
Selecting \Bd and \Bu in a similar phase-space region, 
the ratio of \BdJpsiKst and \BuJpsiK yields (corrected for the ratio of branching ratios) 
is a good measure 
of the ratio of $\epsilon^{\rm REC}$ between \BuJpsiK and \Bqmumu, as shown in Ref.~\cite{roadmap}.   
The measurement from data is $0.59 \pm 0.04$, in good agreement with the estimate from MC simulation ($0.58 \pm 0.02$).

The accuracy of the simulation of $\epsilon^{\rm SEL|REC}$ 
relies on how well the MC describes 
the variables entering the selection. Of these only the \IP distributions show a significant 
discrepancy: the data are measured to have $\sim 10\%$ worse resolution than 
the simulation. 
Smearing the track parameters in MC to reproduce the \IP distribution in data changes 
the selection efficiencies by 5--7\% depending on the channel. 
However, the ratios of $\epsilon_{\rm norm}^{\rm SEL|REC}/\epsilon_{\rm sig}^{\rm SEL|REC}$ 
remain unchanged within the MC statistical uncertainty. 
The ratios predicted by the MC are 
$0.85 \pm 0.01~(\BuJpsiK)$, $0.63 \pm 0.01~(\BsJpsiPhi)$ and $1.09 \pm 0.01~(\BdKpi)$, 
where the uncertainties correspond to the MC statistical uncertainties. The largest contribution to the 
difference in the selection efficiencies for \BuJpsiK and
\BsJpsiPhi compared to the signal comes from the additional $\rm\chi_{IP}^2$ requirements on the extra tracks in 
the normalization channels.
For the \BdKpi normalization channel, the selection efficiency is higher than for the signal 
as the tight ($\pm 60 \MeVcc$) mass window is not applied to the \Bhh channel.  
The ratios of efficiencies including acceptance, reconstruction and selection between normalization and signal decays 
are shown in Table~\ref{tab:norm_summary}.

\subsection{Ratio of trigger efficiencies}
\label{sec:ratio_trig}

The trigger efficiency $\epsilon^{\rm TRIG|SEL}$ can be estimated from data 
as described in Ref.~\cite{roadmap}. Events that would have 
triggered even without the presence of the decay products of the signal 
under study are tagged as TIS events (Trigger Independent of Signal). 
TIS events are mostly triggered by the decay products of the other $b$ which
can be in the acceptance given the forward geometry of LHCb.

If the presence of the signal under study alone is sufficient to 
trigger, events are tagged as TOS 
(Trigger On Signal). An event can also be TIS and TOS simultaneously (TIS\&TOS).
The overall trigger efficiency on selected events can then be expressed as: 
\begin{equation}
\label{eq:etrigsel}
\epsilon^{\rm TRIG|SEL}  =  
\frac{N^{\rm TRIG}}{N^{\rm SEL}} = 
\frac{N^{\rm TIS}}{N^{\rm SEL}} 
\frac{N^{\rm TRIG}}{N^{\rm TIS}} =
\epsilon^{\rm TIS}
\frac{N^{\rm TRIG}}{N^{\rm TIS}} \, ,
\end{equation} 
\noindent where $N^{\rm SEL}$ is not directly observable as it corresponds to a sample of selected events for a  fully efficient trigger.
The TIS efficiency ($\epsilon^{\rm TIS}$) can however be measured directly on data using the ratio  
$ { N^{\rm TIS\&TOS}}/{N^{\rm TOS}}$. Therefore
$\epsilon^{\rm TRIG|SEL}$ can be
expressed in terms of fully observable quantities.

The trigger efficiency for selecting \BuJpsiK and \BsJpsiPhi is obtained from a large inclusive sample of \jpsi events 
using Eq.~\ref{eq:etrigsel}. The result is 
$\epsilon^{\rm TRIG|SEL}_{\jpsi} = (85.9 \pm 0.9_{\rm stat} \pm 2.0_{\rm syst}) \% $, where the systematic uncertainty 
reflects the approximation of the method as seen in the simulation. 
This efficiency is parameterized as a function of the largest $p_{\rm T}$ and the largest \IP of the two 
muons. Using the phase space of the \Bqmumu decay in these two variables, the trigger efficiency for the signal is 
evaluated to be $\epsilon^{\rm TRIG|SEL}_{\Bqmumu} = (89.9 \pm 0.8_{\rm stat} \pm 4.0_{\rm syst}) \% $, where the systematic 
uncertainty is increased to account for the limitations of using only two variables  (the largest $p_T$ and $IP$ of the muons in the final state) to parameterize the trigger response. 

In the case of the \BdKpi normalization channel, the trigger efficiency is computed using the same events that are used 
for the normalization in Eq.~\ref{eq:normAlpha}. Therefore, combining Eqs.~\ref{eq:normAlpha} and \ref{eq:etrigsel} 
results in an expression equivalent to a normalization which uses 
only TIS events. 
The total number of these events after the first trigger steps (L0 and HLT1) is
578, accepting all HLT2 triggers, which does not allow for a precise measurement of $\epsilon^{\rm TIS}$. 
Instead, this efficiency can be measured 
using another control channel, \BuJpsiK, with the result: 
$\epsilon^{\rm TIS}(\textrm{L0} \times \textrm{HLT1}) = (6.9 \pm0.6)\% $.  
The small correction due to the HLT2 trigger inefficiency on selected \BdKpi candidates 
is taken from the trigger emulation.
The ratios $\epsilon_{\rm norm}^{\rm TRIG|SEL}/\epsilon_{\rm sig}^{\rm TRIG|SEL}$ for the three normalization channels are given in Table~\ref{tab:norm_summary}.
   
\subsection{Overall normalization factor}
\label{sec:overall_norm}

The yields needed to evaluate the normalization factor for the two channels containing a $J/\psi$ 
in the final state are obtained from a Gaussian fit to the 
invariant mass distribution. The number of candidates can be seen in 
Table~\ref{tab:norm_summary}, where the uncertainty is dominated by the differences observed using different fitting models.
In the case of the \BdKpi decay, the RICH particle
identification and mass information are used to extract the fraction of $K^+ \pi^{-}$ events 
from the selected inclusive \Bhh sample. 
The efficiency of the kaon and pion identification requirements 
is not needed since 
their ratio is extracted from the known ratio of \Bdpipi and \BdKpi
branching ratios as described in Ref.~\cite{roadmap}. 
The number of TIS \BdKpi events observed is shown in Table~\ref{tab:norm_summary}.
\begin{table}[t]
 \begin{center}
   \caption[]
   {Summary of the factors and their uncertainties needed to calculate the normalization factors ($\alpha_{\Bqmumu}$) 
for the three normalization channels considered. 
The branching ratios are taken from Refs.~\cite{PDG,BELLE_Bs}. 
The trigger efficiency and number of \BdKpi candidates correspond to only TIS events, as described in the text.}
\vspace{3mm}
\footnotesize
   \label{tab:norm_summary}
\begin{tabular}{@{}l@{\hspace{1mm}}c@{\hspace{1mm}}c@{\hspace{1mm}}c@{\hspace{1mm}}c@{\hspace{1mm}}c@{\hspace{1mm}}c@{}}
     \hline  
                  &   \BR     &   $\frac{\rm
\epsilon_{norm}^{REC}
\epsilon_{norm}^{SEL|REC}
}{\rm
\epsilon_{sig}^{REC}
\epsilon_{sig}^{SEL|REC}
}$      &   $\frac{\rm
\epsilon_{norm}^{TRIG|SEL}\TTstrut
}{\rm
\epsilon_{sig}^{TRIG|SEL} \BBstrut
}$  &   
$N_{\rm norm}$ &  $\alpha_{\Bsmumu}$ & $\alpha_{\Bdmumu}$ \\
 & $(\times 10^{-5})$  &  &  &  &  $(\times 10^{-9})$ & $(\times 10^{-9})$  \\
     \hline  
\BuJpsimumuK 
\BBstrut\TTstrut& $5.98 \pm 0.22$ & $0.49 \pm 0.02$ & $0.96 \pm 0.05$ &  
$12,366 \pm 403$ & 
$ 8.4 \pm 1.3$  & $2.27\pm 0.18$   \\
\BsJpsimumuPhiKK \BBstrut& $3.4 \pm 0.9$ & $0.25 \pm 0.02$ & $0.96\pm 0.05$ & 
$760 \pm 71$ & 
$ 10.5 \pm 2.9$  &  $2.83 \pm 0.86 $  \\
\BdKpi \BBstrut& $1.94 \pm 0.06$ & $0.82 \pm 0.06$ & $0.072 \pm 0.010$ &  
$578\pm74$
& $7.3 \pm 1.8$   &  $1.99 \pm 0.40$  \\
     \hline 
   \end{tabular}
 \end{center}
\normalsize 
\end{table}

As can be seen in Table~\ref{tab:norm_summary}, 
the normalization factors calculated using the three complementary channels 
give compatible results. 
The final normalization factor is a weighted average which takes into account
all the sources of correlations, in particular the dominant one coming from the uncertainty
on $f_d/f_s = 3.71 \pm 0.47$ \cite{HFAG}, with the result:

\begin{eqnarray}
\label{eq:alpha_Bs}
\alpha_{\Bsmumu}= (8.6 \pm 1.1) \times 10^{-9}\, , \nonumber\\
\alpha_{\Bdmumu}= (2.24 \pm 0.16) \times 10^{-9}\, .\nonumber
\label{eq:alpha_Bd}
\end{eqnarray} 

\section{Signal and background likelihoods}
\label{sec:likelihood}

After the selection described in Section~\ref{sec:selection} the signal purity 
assuming the SM branching ratio is still about $10^{-3}$ for \Bsmumu and $10^{-4}$ for \Bdmumu.
Further discrimination is achieved
through the
combination of two independent variables: the multivariate analysis discriminant likelihood, \gl,
combining information that is largely based on the topology of the event,
and the invariant mass. The \gl is defined using the statistical method
described in Refs.~\cite{thesis2010_068,Karlen}. The \gl is defined to have a flat distribution  
between zero and one for signal candidates, and to cluster around zero for background candidates.
The geometrical variables included in the definition of the \gl are intended to be a
complete set describing the properties of the decay, and 
the transverse momentum of the \B candidate is 
also included, which is uncorrelated with the invariant mass. The
variables used in the definition of the \gl are:

\begin{itemize}
\item{\it Lifetime of the \B candidate.} This variable is computed using the distance 
between the secondary vertex and primary vertex, and the reconstructed momentum of the \B candidate. When more than one primary vertex is reconstructed, 
the one that gives the minimum \B impact parameter significance is chosen. 
\item{\it Muon impact parameter $\chi^2$.} This is the lowest impact parameter $\chi^2$ of the 
two muon candidates with respect to any primary vertex reconstructed in the event.
\item{\it Impact parameter of the \B candidate.}
\item{\it Distance of closest approach between the two muon candidates.} 
\item{\it Isolation}. For each of the muon candidates, a search is performed for other tracks  
that can make a good vertex with the muon candidate, as in Ref.~\cite{roadmap}. 
The number of compatible tracks is used as the discriminant variable. 
\item{\it Transverse momentum of the \B candidate.}

\end{itemize}
The analysis is performed in two-dimensional bins of invariant mass and GL. 
The invariant mass in the signal regions ($\pm 60$~\mevcc around the \Bs and the \Bd masses) 
is divided into six bins of equal width, and the GL into four bins of equal width distributed 
between zero and one. A probability to be signal or background is assigned to events falling in each bin.

\subsection{Signal geometrical likelihood}
\label{sec:GL}

Although the \gl variable described above was defined using MC events, 
the probability that a signal event has a given value of \gl is obtained from data using inclusive \Bhh events. 
Studies with large samples of MC events show that after reconstruction and selection the \gl distributions obtained 
from \Bqmumu and \Bhh signal events agree within uncertainties ($3\%$). On the other hand, 
the two distributions are different 
after the trigger is emulated. This bias can be removed if only TIS \Bhh events are used in the evaluation 
of the \gl distribution. However, the total 
number of TIS \Bhh events after all trigger steps (L0, HLT1 and HLT2) is
152 which is insufficient.
Instead, for the \Bhh events, the first two trigger steps (L0 and HLT1) are required to be TIS while 
at the HLT2 step any of the HLT2 triggers are accepted. This yields 955 events.   
The \gl distribution obtained using these events is corrected for the small bias ($< 5\%$) 
introduced at the HLT2 stage using the trigger emulation.
Detailed checks with a large sample of \DKpi decays have validated this 
procedure.

\begin{figure}[htbp]
  \begin{center}
  \ifthenelse{\boolean{pdflatex}}{
     \includegraphics*[width=0.7\textwidth]{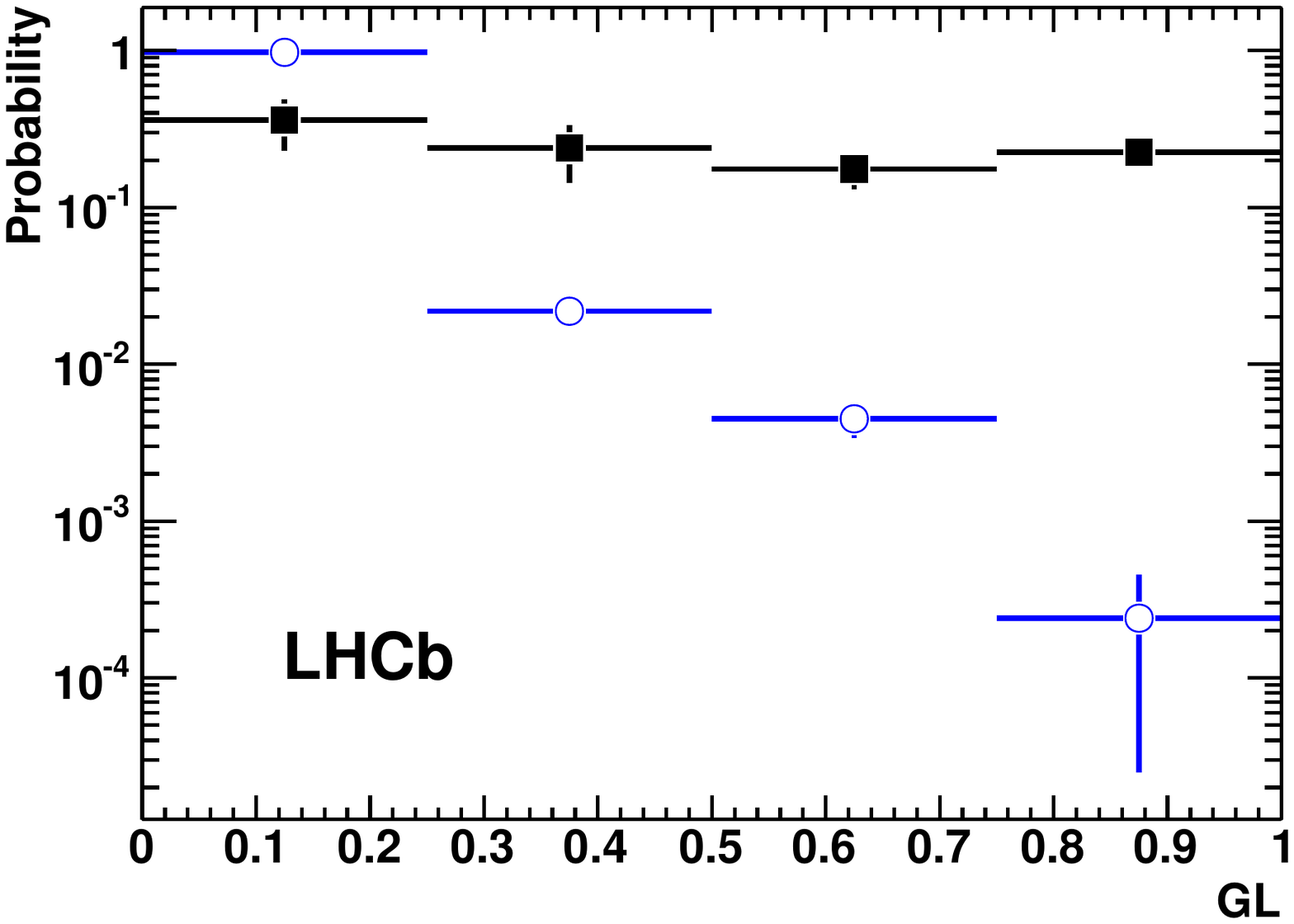}
   }{
    \includegraphics*[width=0.7\textwidth]{GL_pdfs_sig_bkg.eps}
   }
  \end{center}
  \caption{Probability of signal events in bins of GL obtained from the inclusive sample of TIS \Bhh events (solid squares). 
The background probability (open circles) is obtained from the events in the sidebands of the $\mu\mu$ 
invariant mass distribution in the \Bs mass window.}
\label{fig:GL_data}
\end{figure}

The number of TIS \Bhh events in each \gl bin is obtained from a fit
to the inclusive mass distribution~\cite{roadmap2} assigning the muon
mass to the two particles. 
The measured fractions in each \gl bin can be seen in Fig.~\ref{fig:GL_data} 
and are quoted in Table~\ref{tab:gl_pdf}. 
The systematic uncertainties are included, estimated by comparing the results
from the inclusive \Bhh fit model with those obtained using a double Crystal Ball function~\cite{crystalball} and 
a simple background subtraction. 
The measured \gl 
distribution obtained from TIS \Bhh events is compatible with a flat distribution, as expected if the simulation 
reproduces correctly the data. 

\subsection{Signal invariant mass likelihood}
\label{sec:mass}

The signal mass lineshape is parameterized using a Crystal Ball function~\cite{crystalball}.
Two methods have been used to estimate the \Bqmumu mass resolution from data. 
The first of these methods uses 
an interpolation between the measured resolutions for $c\overline{c}$ resonances (\jpsi, \psitwos) and
$b\overline{b}$ resonances (\OneS, \TwoS, \ThreeS) decaying into two muons. 
It has been observed that over 
this mass range the dimuon invariant mass resolution depends linearly on the invariant mass 
of the muon pair to 
good approximation. Events selected in the mass ranges around the $c\overline{c}$ and 
$b\overline{b}$ resonances were weighted such that the momentum spectra of these resonances reproduce the 
expected momentum spectrum of the $b$ hadron in the decay \Bqmumu. The mass resolutions 
of the $c\overline{c}$ and 
$b\overline{b}$ resonances were then determined fitting a Crystal Ball (\jpsi, \OneS) or
a Gaussian (\psitwos, \TwoS and \ThreeS) over exponential backgrounds. 

The mass resolution is defined as the $\sigma$ of the Crystal Ball when there is sufficient data 
to perform a fit with the Crystal Ball function (\jpsi and \OneS). Otherwise
a Gaussian fit is made (\psitwos, \TwoS, \ThreeS) and
the $\sigma$ of the Gaussian is used as an estimator of the $\sigma$ of the Crystal Ball.
For the Crystal Ball function, the parameters describing the radiative tail are 
in good agreement between data and the Monte Carlo simulation. 
No systematic shifts in the resolution has been found by using 
a Crystal Ball or a Gaussian above the transition point.

Interpolating linearly between the five fitted resolutions to $M_{\Bs}$ an invariant mass resolution 
of $\sigma = 26.83 \pm  0.14\,\mevcc$ was estimated for \Bqmumu.
The systematic 
uncertainty is estimated to be 1\mevcc mainly due to the reweighting of the momentum spectrum of the dimuon
resonances and the variation of the resolution over the width of the \Bqmumu signal region.

\begin{figure}[t]
  \begin{center}
  \ifthenelse{\boolean{pdflatex}}{
     \includegraphics*[width=0.7\textwidth]{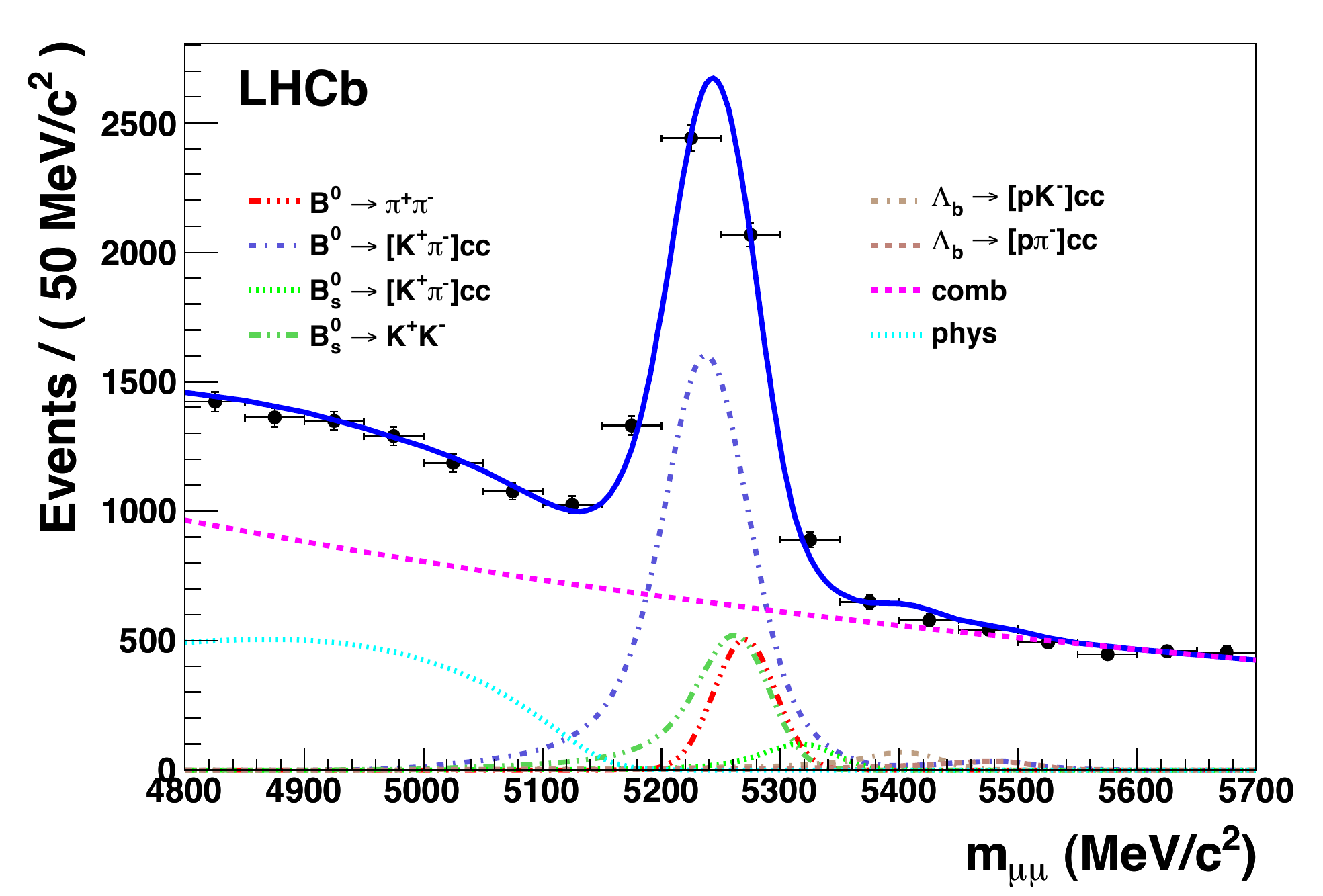}
   }{
    \includegraphics*[width=0.7\textwidth]{Mass_B2hh.eps}
   }
  \end{center}
\vspace*{-5mm}
\caption{Fit of the invariant mass distribution for \Bhh candidates in the \gl range [0.25, 1.0]. 
The pink dashed curve is the combinatorial background component, while the physical background is shown 
with a light-blue dotted curve. The remaining contributions are from the two-body decays of the \Bd, \Bs and \Lambdab.}
\label{fig:mass_b2hh}
\end{figure}

The second method that was used to estimate the invariant mass resolution from data
is to use the inclusive \Bhh sample. The particle identification requirement 
would modify the momentum and transverse momentum spectrum of pions and kaons, 
and thus the mass resolution. Therefore, the fit is performed to the
inclusive \Bhh sample without requiring particle identification and 
assigning the muon mass to the decay products. 
The fit has been performed in the \gl range [0.25,1.0]
and the results are shown in Fig.~\ref{fig:mass_b2hh}.  The fitted parameters
are: the mass resolution, the \Bd and \Bs masses, the signal yield, the
combinatorial background yields, as well as the fraction of radiative tail and
the parameters that describe the combinatorial background. The relative contributions of \Bd and \Bs 
decays are fixed to their known values. The result of the
fit for the mass resolution, $\sigma = 25.8 \pm 1.0 \,\mevcc$, is consistent with the value
obtained from the interpolation method. However, by varying the assumptions
made for the parameters describing the partially reconstructed three-body 
$b$-hadron decays (physical background), the estimate obtained for
the resolution can change by up to 2.7\mevcc. This is assigned as systematic
uncertainty for this method.

The weighted average of the two methods, $\sigma = 26.7 \pm 0.9 \mevcc$, is taken as the
invariant mass resolution and considered to be the same for \Bd and \Bs
decays. 
The mean values of the masses 
obtained from the inclusive \Bhh fit are consistent with, but not as precise as, the
values obtained using the exclusive decay modes \BdKpi and \BsKK isolated using the RICH particle 
identification: $M_{\Bd} =
5275.0 \pm 1.0 \,\mevcc$ and $M_{\Bs} = 5363.1 \pm 1.5 \,\mevcc$, which are used in
the evaluation of the invariant mass likelihood. The mean values of the masses are $\sim 0.07 \%$ 
below the known values~\cite{PDG} which is attributed to a small residual miscalibration of the
magnetic field map. However this has no impact on the analysis, provided that
the search windows are centred around the measured values.

\subsection{Background likelihood}
\label{sec:background}
\begin{table}
\caption{
Probability of signal events in bins of GL obtained from the inclusive sample of TIS \Bhh events. 
The background probability in the \Bs mass window is obtained from the events in the sidebands 
of the dimuon invariant mass distribution. 
}
\begin{center}
\begin{tabular}{ccc}
\hline 
\ GL bin & Signal prob. & Background prob. \\ 
\hline 
\ $0.0 - 0.25$ & $0.360 \pm 0.130$ &  $0.9735^{+0.0030}_{-0.0032}$ \\ 
\ $0.25 - 0.5$ & $0.239 \pm 0.096$ &  $0.0218^{+0.0030}_{-0.0028}$ \\ 
\ $0.5 - 0.75$ & $0.176 \pm 0.046$ &  $0.0045^{+0.0012}_{-0.0010}$ \\ 
\ $0.75 - 1.0$ & $0.225 \pm 0.036$ &  $0.00024^{+0.00031}_{-0.00015}$ \\ 
\hline 
\end{tabular}
\end{center}
\label{tab:gl_pdf}
\end{table}

\begin{figure}[tbp]
\centering
\begin{minipage}[b]{0.49\linewidth}
\includegraphics[width=\textwidth]{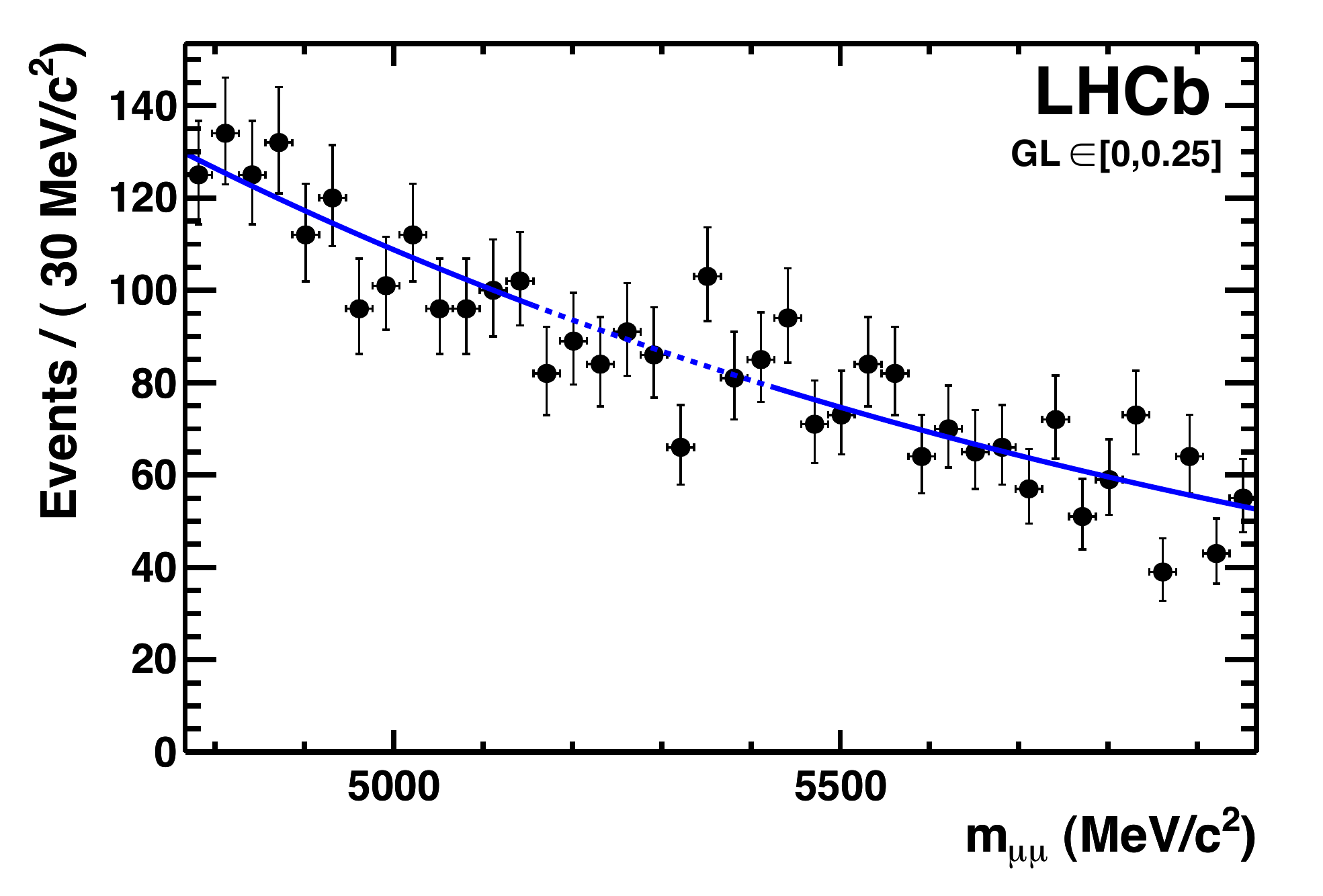}
\hspace{-75mm}\textbf{(a)}
\end{minipage}
\begin{minipage}[b]{0.49\linewidth}
\includegraphics[width=\textwidth]{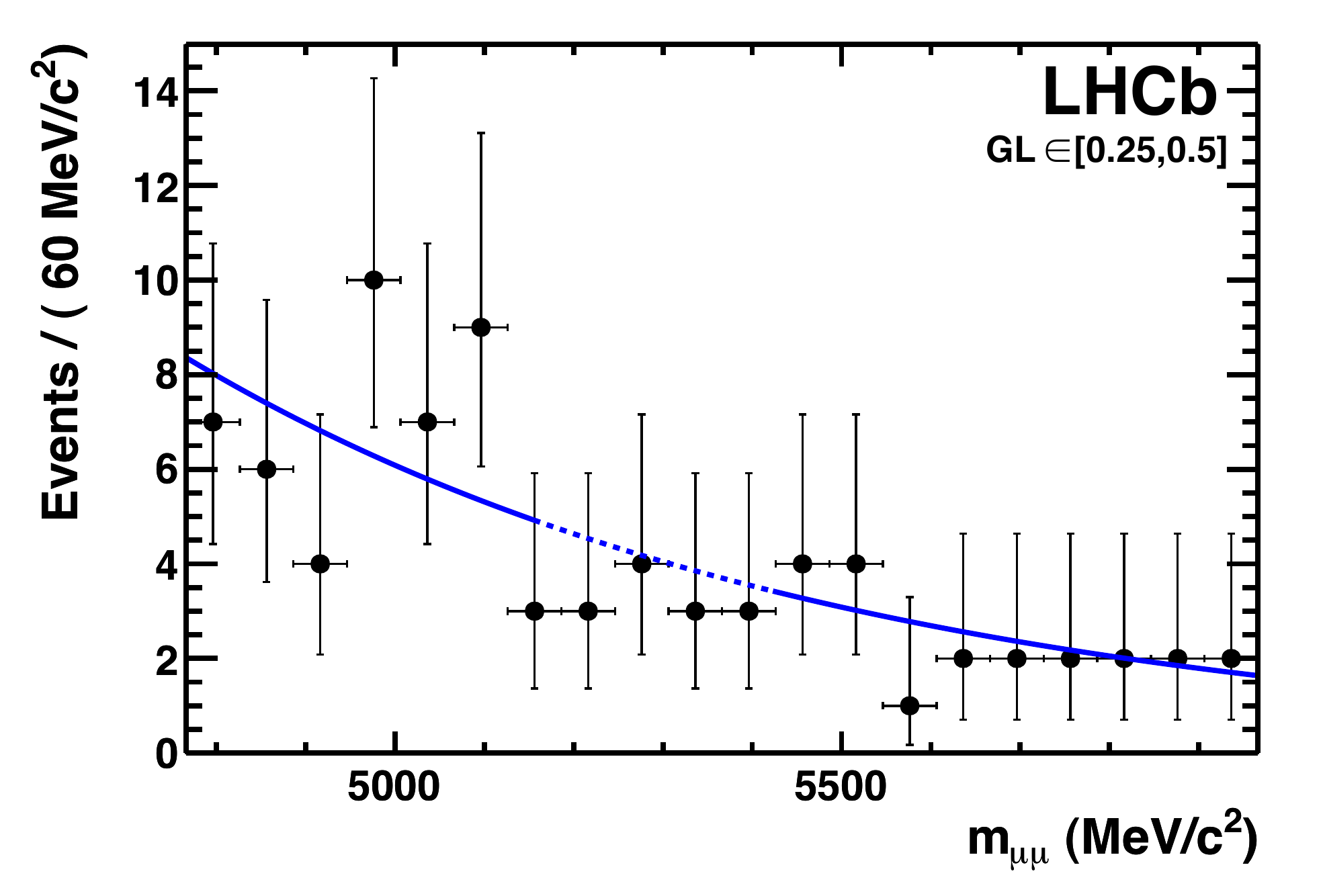}
\hspace{-75mm}\textbf{(b)}
\end{minipage}
\begin{minipage}[b]{0.49\linewidth}
\includegraphics[width=\textwidth]{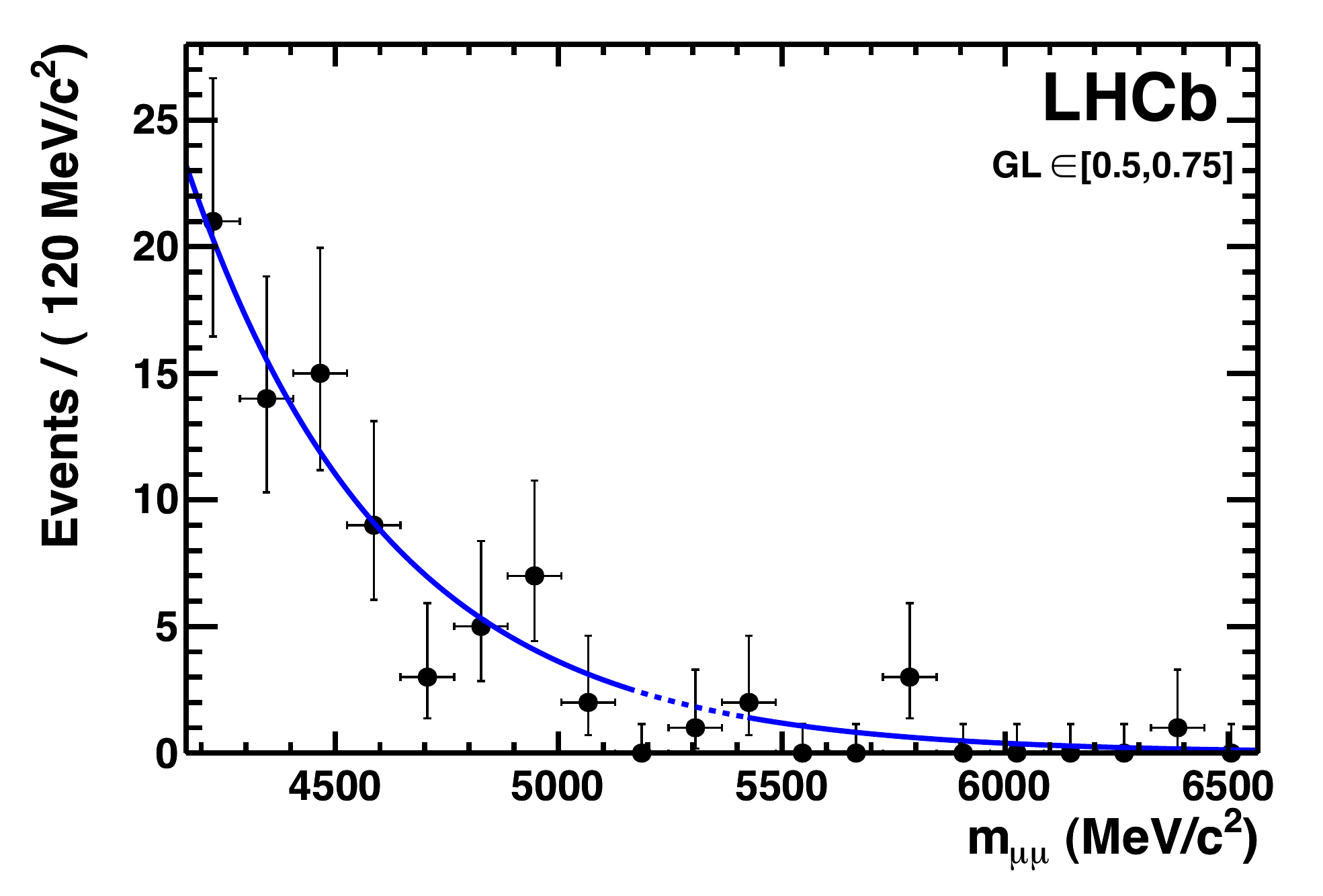}
\hspace{-75mm}\textbf{(c)}
\end{minipage}
\begin{minipage}[b]{0.49\linewidth}
\includegraphics[width=\textwidth]{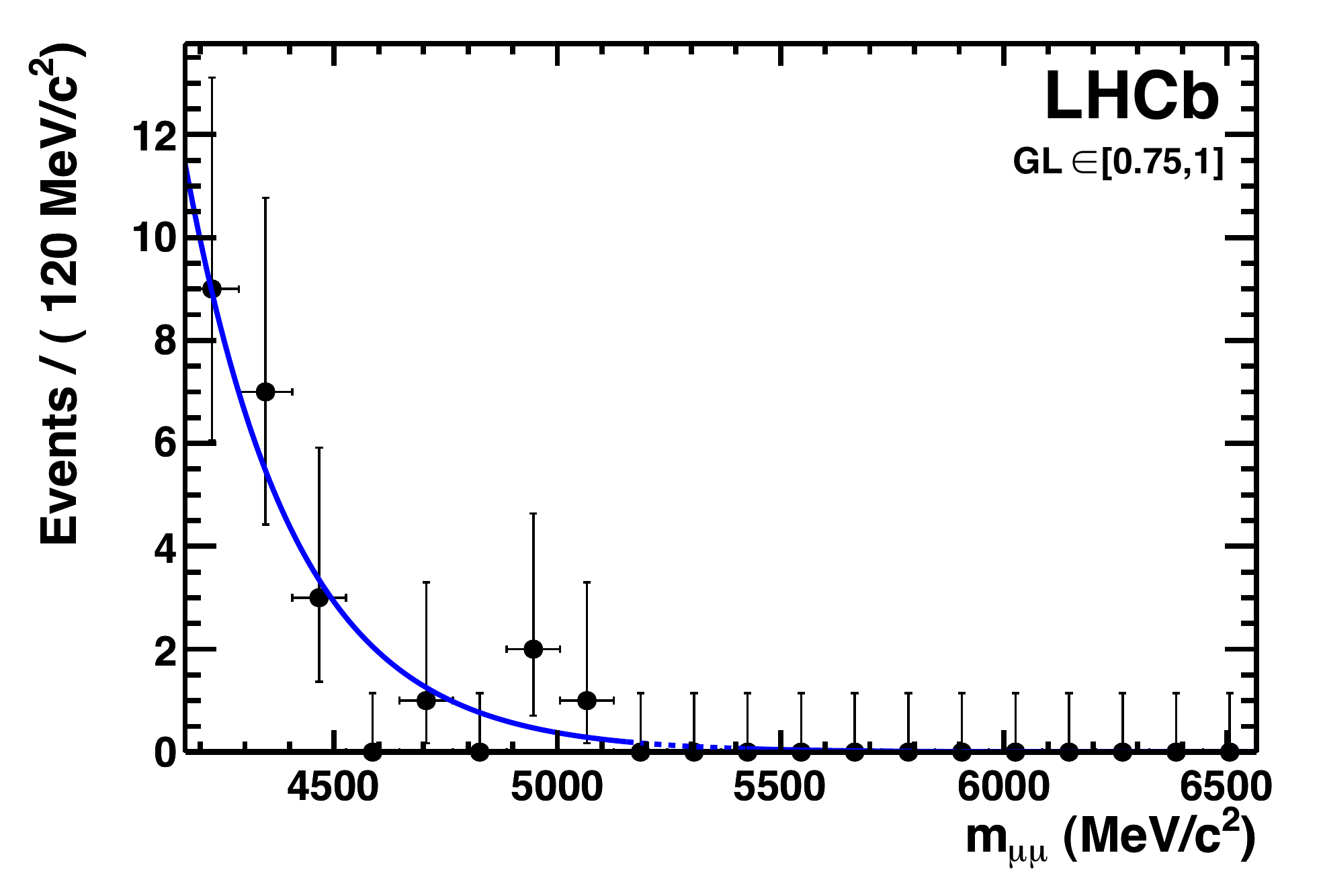}
\hspace{-75mm}\textbf{(d)}
\end{minipage}
%
\caption{ Distribution of the $\mu\mu$ invariant mass for different \gl bins: (a) 
$[0,0.25]$, (b) $[0.25,0.5]$, (c) $[0.5,0.75]$, (d) $[0.75,1.0]$. 
The blue solid lines show the interpolation model used and
the dashed line shows the result of the interpolation in the search windows.}
\label{fig:mumu_bkg_sidebands}
\end{figure}

The mass sidebands are defined in the range between $M_{\Bq} \pm 600~(1200) \mevcc$ for the 
lower (upper) two \gl bins, excluding the two search windows ($M_{\Bq} \pm 60 \mevcc$).
The background in the mass sidebands is fitted with an exponential function, $f(M) = A e^{- k M}$.
The value of the exponential index $k$ is fitted independently in each \gl bin, in order 
to account for potentially different background compositions. The
distribution of the invariant mass for each 
\gl bin is shown in Fig.~\ref{fig:mumu_bkg_sidebands}, and the predictions for the numbers of events 
in the signal regions can be seen 
in Tables~\ref{tab:data_bsmm} and \ref{tab:data_bdmm}. 
The background probability in the \Bs mass window
as a function of \gl is shown in Fig.~\ref{fig:GL_data} and in Table~\ref{tab:gl_pdf}.
The results have been checked by fixing the exponential index $k$ to be the same 
in all \gl bins, using a double exponential, or using a simple linear fit in the
region around the signal window. In all cases the 
predicted background is consistent with the result of the exponential fit with different $k$ values, although the quality of the fit is significantly worse when $k$ is forced to be the same for all bins.

\section{Results}
\label{sec:results}

\begin{figure}[t]
  \begin{center}
  \ifthenelse{\boolean{pdflatex}}{
     \includegraphics*[width=0.7\textwidth]{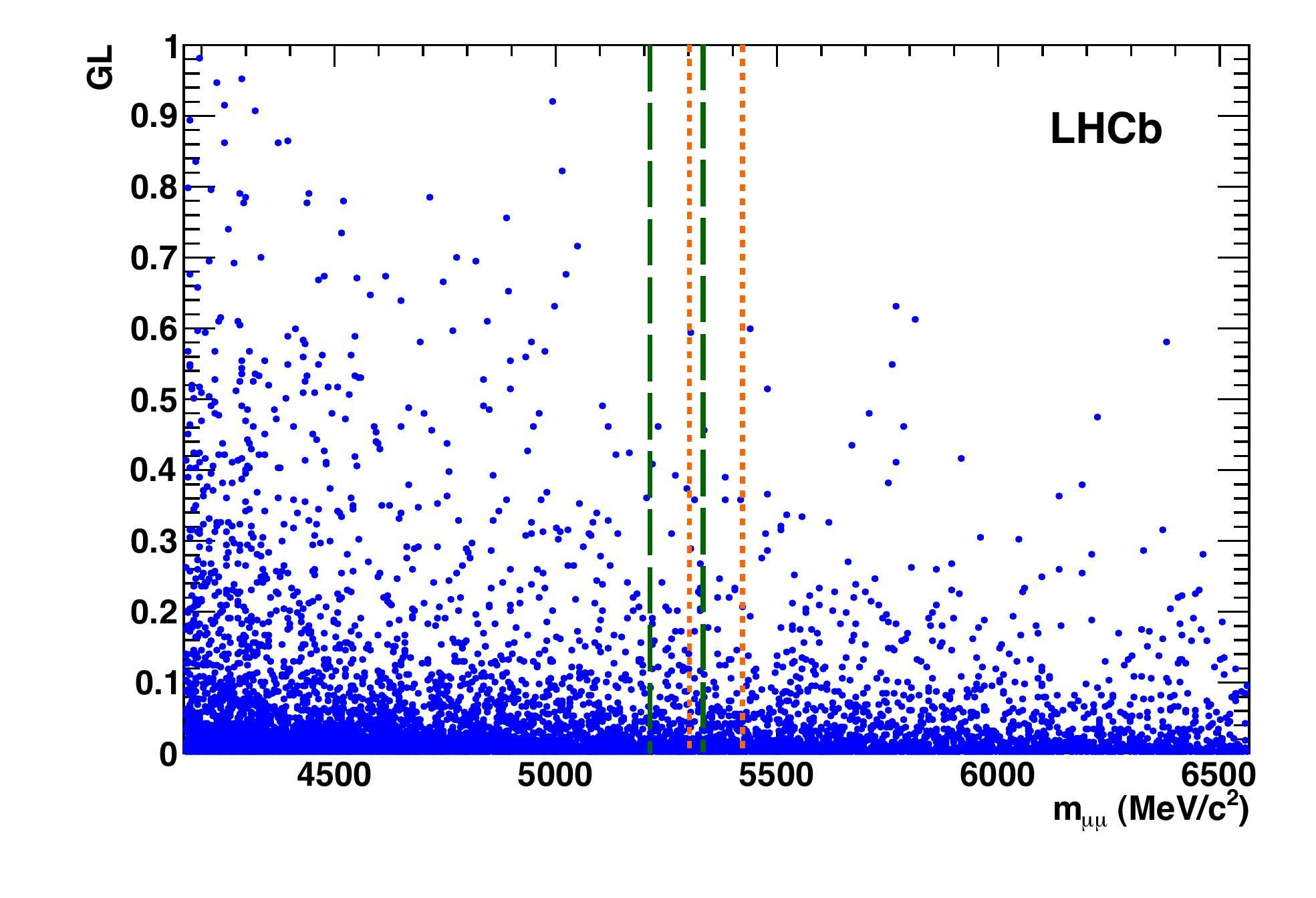}
   }{
     \includegraphics*[width=0.7\textwidth]{GLvsmass.eps}
   }
  \end{center}
  \caption{Observed distribution of selected dimuon events in the GL
    {\it vs} invariant mass plane. The orange short-dashed 
(green long-dashed) lines indicate the $\pm 60 \mevcc$ search window around
    the \Bs (\Bd).}
\label{fig:GLvsmass_obs}
\end{figure}

\begin{figure}[tbp]
\centering
\begin{minipage}[b]{0.49\linewidth}
\includegraphics[width=\textwidth]{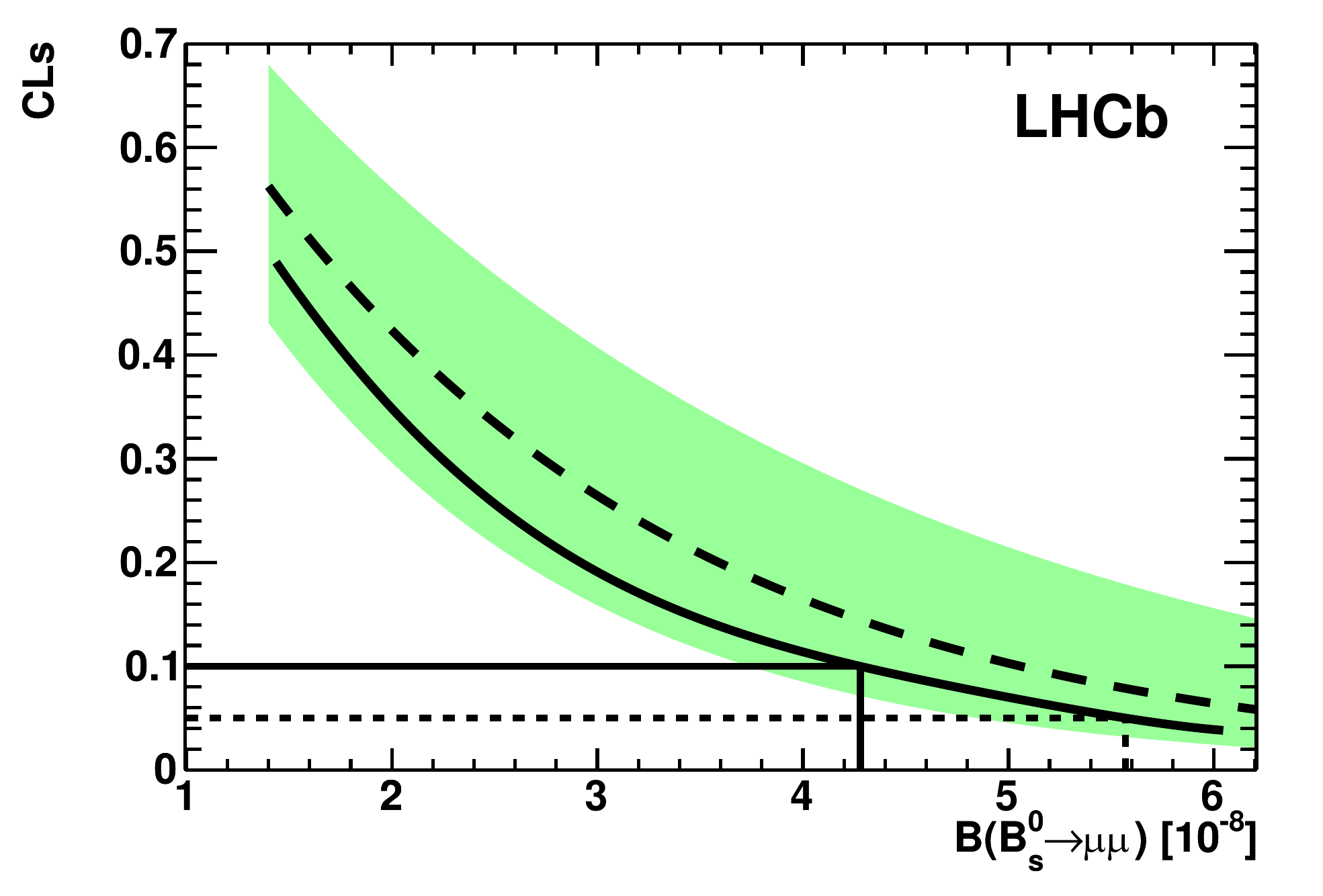}
\hspace{-75mm}\textbf{(a)}
\end{minipage}
\begin{minipage}[b]{0.49\linewidth}
\includegraphics[width=\textwidth]{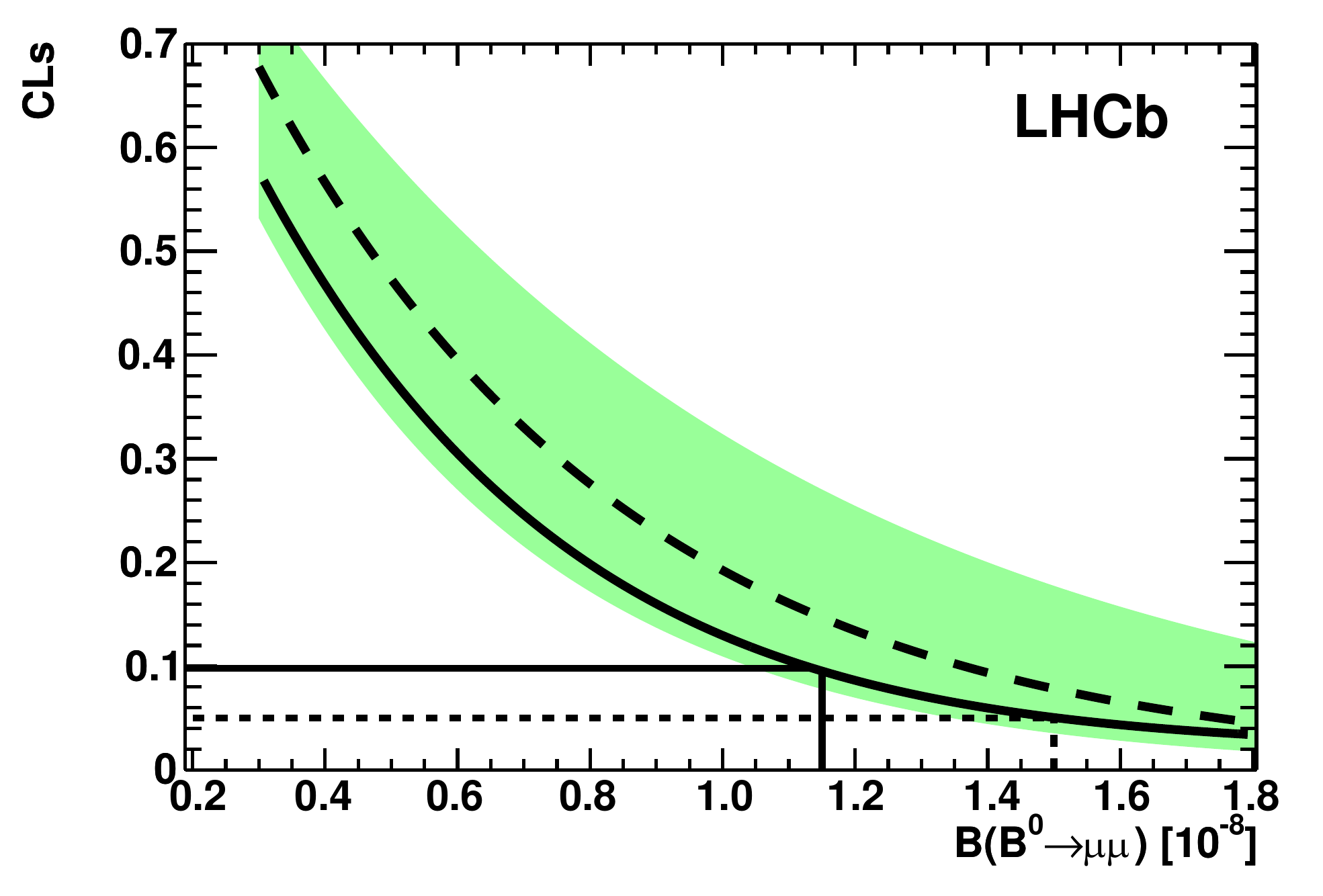}
\hspace{-75mm}\textbf{(b)}
\end{minipage}
\caption
{(a) Observed (solid curve) and expected (dashed curve) \CLs values as a function of \BRof{\Bsmm}.
The green shaded area contains the $\pm 1 \sigma$ interval of possible results compatible with the expected value when only background
is observed. The 90\,\% (95\,\%) \CL  observed value is identified by the solid (dashed) line. 
(b) the same for \BRof{\Bdmm}.
}
\label{fig:CLsvsBR}
\end{figure}

For each of the 24 bins (4 bins in \gl and 6 bins in mass) the
expected number of background events is computed from the fits to the invariant
mass sidebands described in Section~\ref{sec:background}. The results are
shown in Tables~\ref{tab:data_bsmm} and \ref{tab:data_bdmm}. The expected
numbers of signal events are computed using the normalization factors from Section~\ref{sec:normalization},
and the signal likelihoods computed in 
Section~\ref{sec:GL} and Section~\ref{sec:mass} for a given value of
\BRof{\Bqmumu}. The expected numbers of signal events for the
SM branching ratios are shown in Tables~\ref{tab:data_bsmm} and
\ref{tab:data_bdmm}. The distribution of observed events in the GL {\it vs} invariant mass 
plane can be seen in Fig.~\ref{fig:GLvsmass_obs}, and the observed number
of events in each bin are given in Tables~\ref{tab:data_bsmm} and
\ref{tab:data_bdmm}.  
The compatibility of the observed distribution of events in the GL {\it vs} invariant mass plane 
with a given branching ratio hypothesis is evaluated using the \CLs method~\cite{Read_02}. 
This provides two estimators: \CLs is a measure of the compatibility of the observed distribution with the 
signal hypothesis, 
while \CLb is a measure of the compatibility with the 
background-only hypothesis. 
The observed distribution
of \CLs as a function of the assumed branching ratio can be seen in Fig.~\ref{fig:CLsvsBR}. The 
expected distributions of possible values of \CLs assuming the background-only hypothesis are also shown in the same 
figure as a green shaded area that covers the region of $\pm 1 \sigma$ of background compatible observations. 
The uncertainties in the signal and background likelihoods (Section~\ref{sec:likelihood}) and 
normalization factors (Section~\ref{sec:normalization}) are used to compute the uncertainties in the background 
and signal predictions in Tables~\ref{tab:data_bsmm} and \ref{tab:data_bdmm}. These 
uncertainties are the only source of systematic uncertainty and they are included 
in the \CLs using the techniques described in Ref.~\cite{Read_02}.
Given the specific pattern of the observed events,
the systematic uncertainty on the background prediction has a negligible effect on the 
quoted limit. The effect of the uncertainty on the signal prediction increases the quoted 
limits by less than $3\%$. 

The evaluation of \CLb~\cite{Read_02} 
gives a probability of about $20\%$ 
for the compatibility with the 
background-only hypothesis for both the \Bs and \Bd decays. 
This low value can be attributed to the slight deficit of observed 
events in the most sensitive bins, as can be seen in Tables~\ref{tab:data_bsmm} and \ref{tab:data_bdmm}.
As no significant deviation from the background-only hypothesis 
is observed, upper limits are computed using the \CLs distributions in Fig.~\ref{fig:CLsvsBR} with the results
\begin{eqnarray}
\BRof{\Bsmm} &<& 4.3~(5.6)\times10^{-8}~{\rm at}~90\,\% ~(95\,\%)~{\rm C.L.,}  \nonumber \\
\BRof{\Bdmm} &<& 1.2~(1.5)\times10^{-8}~{\rm at}~90\,\% ~(95\,\%)~{\rm C.L.,} \nonumber
\end{eqnarray} 
while the expected values of the limits are $\BRof{\Bsmm} <
5.1~(6.5)\times10^{-8}$ 
and $\BRof{\Bdmm} < 1.4~(1.8)\times10^{-8}~{\rm at}~90\,\% ~(95\,\%)$ \CL 
The limits observed are similar to the best 
published limits~\cite{d0_PLB} for the decay \Bsmumu and more restrictive for the decay \Bdmumu~\cite{cdf_PRL}.

\begin{table}[t]
\caption{Expected background, expected SM signal and observed number of events
  in bins of \gl and invariant mass, in the $\pm 60 \mevcc$ mass window around
  the \Bs mass central value of 5363.1 \MeVcc.}
\begin{center}
\resizebox {\textwidth }{!}{%
\begin{tabular}{|l c l|c|c|c|c|}
\hline
& & & \multicolumn{4}{c|}{\gl bin\TTstrut\BBstrut}  \\
& & & [0, 0.25]  & [0.25, 0.5] & [0.5, 0.75] & [0.75, 1] \\
\hline
\multirow{22}{*}{\rotatebox{90}{Invariant mass bin (\MeVcc)}}
         & \multirow{3}{*}{[$-60$, $-40$]}  
                 &  Exp.\ bkg.& $56.9^{+1.1}_{-1.1}$ & $1.31^{+0.19}_{-0.17}$ & $0.282^{+ 0.076}_{-0.065}$ & $0.016^{+0.021}_{-0.010}$ \TTstrut\\
                 && Exp.\ sig.
                              & $0.0076^{+0.0034}_{-0.0030}$ & $0.0050^{+0.0027}_{-0.0020}$ & $0.0037^{+0.0015}_{-0.0011}$ & $0.0047^{+0.0015}_{-0.0010}$ \TTstrut\\
                 && Observed     & $39$               & $2$                 & $1$                     & $0$  \\
\cline{2-7}
         & \multirow{3}{*}{[$-40$, $-20$]}  
                 &  Exp.\ bkg.& $56.1^{+1.1}_{-1.1}$ & $1.28^{+0.18}_{-0.17}$ & $0.269^{+ 0.072}_{-0.062}$ & $0.0151^{+0.0195}_{-0.0094}$ \TTstrut\\
                 && Exp.\ sig.
                               & $0.0220^{+0.0084}_{-0.0081}$ & $0.0146^{+0.0067}_{-0.0054}$ & $0.0107^{+0.0036}_{-0.0027}$ & $0.0138^{+0.0035}_{-0.0025}$ \TTstrut\\

                 && Observed     & $55$ & $2$ & $0$ & $0$  \Bstrut\\
\cline{2-7}
         & \multirow{3}{*}{[$-20$, 0]}  
                 &  Exp.\ bkg.& $55.3^{+1.1}_{-1.1}$ & $1.24^{+0.17}_{-0.16}$ & $0.257^{+ 0.069}_{-0.059}$ & $0.0139^{+0.0179}_{-0.0086}$ \TTstrut\\
                 && Exp.\ sig.
                              & $0.038^{+0.015}_{-0.015}$ & $0.025^{+0.012}_{-0.010}$ & $0.0183^{+0.0063}_{-0.0047}$ & $0.0235^{+0.0060}_{-0.0044}$ \TTstrut\\

                 && Observed     & $73$ & $0$ & $0$ & $0$  \Bstrut\\
\cline{2-7}
         & \multirow{3}{*}{[0, 20]}   
                 &  Exp.\ bkg.& $54.4^{+1.1}_{-1.1}$ & $1.21^{+0.17}_{-0.16}$ & $0.246^{+ 0.066}_{-0.057}$ & $0.0128^{+0.0165}_{-0.0080}$ \TTstrut\\
                 && Exp.\ sig.
                              & $0.038^{+0.015}_{-0.015}$ & $0.025^{+0.012}_{-0.010}$ & $0.0183^{+0.0063}_{-0.0047}$ & $0.0235^{+0.0060}_{-0.0044}$ \TTstrut\\

                 && Observed     & $60$ & $0$ & $0$ & $0$  \Bstrut\\
\cline{2-7}
         & \multirow{3}{*}{[20, 40]}   
                 &  Exp.\ bkg.& $53.6^{+1.1}_{-1.0}$ & $1.18^{+0.17}_{-0.15}$ & $0.235^{+ 0.063}_{-0.054}$ & $0.0118^{+0.0152}_{-0.0073}$ \TTstrut\\
                 && Exp.\ sig.
                              & $0.0220^{+0.0084}_{-0.0081}$ & $0.0146^{+0.0067}_{-0.0054}$ & $0.0107^{+0.0036}_{-0.0027}$ & $0.0138^{+0.0035}_{-0.0025}$ \TTstrut\\

                 && Observed     & $53$ & $2$ & $0$ & $0$  \Bstrut\\
\cline{2-7}
         & \multirow{3}{*}{[40, 60]}   
                 &  Exp.\ bkg.& $52.8^{+1.0}_{-1.0}$ & $1.14^{+0.16}_{-0.15}$ & $0.224^{+ 0.060}_{-0.052}$ & $0.0108^{+0.0140}_{-0.0068}$ \TTstrut\\
                 && Exp.\ sig.
                              & $0.0076^{+0.0031}_{-0.0027}$ & $0.0050^{+0.0025}_{-0.0019}$ & $0.0037^{+0.0013}_{-0.0010}$ & $0.0047^{+0.0013}_{-0.0010}$ \TTstrut\\

                 && Observed     & $55$ & $1$ & $0$ & $0$  \BBstrut\\

\hline
\end{tabular}
}
\end{center}
\label{tab:data_bsmm}
\end{table}

\begin{table}[t]
\caption{Expected background, expected SM signal and observed number of events
  in bins of \gl and invariant mass, in the $\pm 60 \mevcc$ mass window around
  the \Bd central value of 5275.0 \MeVcc.}
\begin{center}
\resizebox {\textwidth }{!}{%
\begin{tabular}{|l c l|c|c|c|c|}
\hline
& & & \multicolumn{4}{c|}{\gl bin\TTstrut\BBstrut}  \\
& & & [0, 0.25] & [0.25, 0.5] & [0.5, 0.75] & [0.75, 1] \\
\hline
\multirow{22}{*}{\rotatebox{90}{Invariant mass bin (\MeVcc)}} 
        & \multirow{3}{*}{[$-60$, $-40$]}    
             &  Exp.\ bkg.& $60.8^{+1.2}_{-1.1}$ & $1.48^{+0.19}_{-0.18}$ & $0.345^{+ 0.084}_{-0.073}$ & $0.024^{+0.027}_{-0.014}$ \TTstrut\\
             && Exp.\ sig.
                          & $0.00090^{+0.00036}_{-0.00035}$ & $0.00060^{+0.00029}_{-0.00023}$ & $0.00044^{+ 0.00016}_{-0.00012}$ & $0.00056^{+0.00015}_{-0.00011}$ \TTstrut\\
             && Observed     & $59$ & $2$ & $0$ & $0$ \Bstrut \\ 
\cline{2-7}          
        &  \multirow{3}{*}{[$-40$, $-20$]}    
             &  Exp.\ bkg.& $59.9^{+1.1}_{-1.1}$ & $1.44^{+0.19}_{-0.17}$ & $0.329^{+ 0.080}_{-0.070}$ & $0.022^{+0.024}_{-0.013}$ \TTstrut\\
             && Exp.\ sig.
                          & $0.00263^{+0.00093}_{-0.00093}$ & $0.00174^{+0.00076}_{-0.00061}$ & $0.00128^{+ 0.00038}_{-0.00030}$ & $0.00164^{+0.00035}_{-0.00025}$ \TTstrut\\
             && Observed     & $67$ & $0$ & $0$ & $0$ \Bstrut \\ 
\cline{2-7}         
        &  \multirow{3}{*}{[$-20$, 0]}    
             &  Exp.\ bkg.& $59.0^{+1.1}_{-1.1}$ & $1.40^{+0.18}_{-0.17}$ & $0.315^{+ 0.077}_{-0.067}$ & $0.020^{+0.022}_{-0.012}$ \TTstrut\\
             && Exp.\ sig.& $0.0045^{+0.0017}_{-0.0017}$ & $0.0030^{+0.0014}_{-0.0011}$ & $0.00219^{+ 0.00067}_{-0.00054}$ & $0.00280^{+0.00060}_{-0.00045}$ \TTstrut\\
             && Observed     & $56$ & $2$ & $0$ & $0$ \Bstrut \\
\cline{2-7}          
        &  \multirow{3}{*}{[0, 20]}    
             &  Exp.\ bkg.& $58.1^{+1.1}_{-1.1}$ & $1.36^{+0.18}_{-0.16}$ & $0.300^{+ 0.073}_{-0.064}$ & $0.019^{+0.021}_{-0.011}$ \TTstrut\\
             && Exp.\ sig.& $0.0045^{+0.0017}_{-0.0017}$ & $0.0030^{+0.0014}_{-0.0011}$ & $0.00219^{+ 0.00067}_{-0.00054}$ & $0.00280^{+0.00060}_{-0.00045}$ \TTstrut\\
             && Observed     & $60$ & $0$ & $0$ & $0$ \Bstrut \\ 
\cline{2-7}         
        &  \multirow{3}{*}{[20, 40]}    
             &  Exp.\ bkg.& $57.3^{+1.1}_{-1.1}$ & $1.33^{+0.17}_{-0.16}$ & $0.287^{+ 0.070}_{-0.061}$ & $0.017^{+0.019}_{-0.010}$ \TTstrut\\
             && Exp.\ sig.
                          & $0.00263^{+0.00093}_{-0.00093}$ & $0.00174^{+0.00076}_{-0.00061}$ & $0.00128^{+ 0.00038}_{-0.00030}$ & $0.00164^{+0.00035}_{-0.00025}$ \TTstrut\\
             && Observed     & $42$ & $2$ & $1$ & $0$ \Bstrut \\  
\cline{2-7}        
        &  \multirow{3}{*}{[40, 60]}    
             &  Exp.\ bkg.& $56.4^{+1.1}_{-1.1}$ & $1.29^{+0.17}_{-0.16}$ & $0.274^{+ 0.067}_{-0.058}$ & $0.0158^{+0.0175}_{-0.0094}$ \TTstrut\BBstrut\\
             && Exp.\ sig.
                          & $0.00090^{+0.00033}_{-0.00032}$ & $0.00060^{+0.00027}_{-0.00021}$ & $0.00044^{+ 0.00014}_{-0.00011}$ & $0.00056^{+0.00013}_{-0.00010}$ \TTstrut\\
             && Observed     & $49$ & $2$ & $0$ & $0$ \BBstrut \\
\hline
\end{tabular}
}
\end{center}
\label{tab:data_bdmm}
\end{table}

\section{Conclusions}
\label{sec:conclusions}

With about 37\invpb of integrated luminosity, LHCb has searched for the rare decays \Bsmumu and \Bdmumu and reached sensitivities 
similar to the existing limits from the Tevatron. This could be achieved
due to the large acceptance and trigger efficiency of LHCb, as well as the larger  
$b \bar{b}$ cross-section in $pp$ collisions at $\sqrt{s}$ = 7~TeV.
The observed events are compatible with the background expectations, and the 
upper limits are evaluated to be
\begin{eqnarray}
\BRof{\Bsmm} &<& 5.6 \times10^{-8}~{\rm at}~95\,\%~{\rm C.L.,}  \nonumber \\
\BRof{\Bdmm} &<& 1.5 \times10^{-8}~{\rm at}~95\,\%~{\rm C.L.,}  \nonumber
\end{eqnarray} 
while the expected values of the limits are $\BRof{\Bsmm} < 6.5\times10^{-8}$ and 
$\BRof{\Bdmm} < 1.8\times10^{-8}~{\rm at}~95\,\%$ C.L. 

The LHC is expected to deliver a much larger sample of $pp$ collisions in 2011.
Given the low level of background in the most sensitive bins shown in 
Tables~\ref{tab:data_bsmm} and \ref{tab:data_bdmm}, LHCb should be able to
explore the interesting region of branching ratios at the $10^{-8}$ level in the near future.

\section*{Acknowledgments}

We express our gratitude to our colleagues in the CERN accelerator departments for the excellent 
performance of the LHC. We thank the technical and administrative staff at CERN and at the LHCb institutes, 
and acknowledge support from the National Agencies: CAPES, CNPq, FAPERJ and FINEP (Brazil); CERN; 
NSFC (China); CNRS/IN2P3 (France); BMBF, DFG, HGF and MPG (Germany); SFI (Ireland); INFN (Italy); 
FOM and NWO (Netherlands); SCSR (Poland); ANCS (Romania); MinES of Russia and Rosatom (Russia); 
MICINN, XUNGAL and GENCAT (Spain); SNSF and SER (Switzerland); NAS Ukraine (Ukraine); 
STFC (United Kingdom); NSF (USA). We also acknowledge the support received from the ERC under FP7 
and the R\'egion Auvergne.

\end{document}